\documentclass[12pt]{article}
\usepackage{amsmath,amssymb}
\usepackage{bm}
\usepackage{color}
\usepackage{cite}
\usepackage{mathtools}

\usepackage{multirow}

\usepackage{dsfont}

\usepackage{here}

\begin{document}
\title{
\begin{flushright}
\ \\*[-80pt]
\begin{minipage}{0.2\linewidth}
\normalsize
\end{minipage}
\end{flushright}
{\Large \bf
Modular flavor symmetric models}
\footnote{The contribution to a special book dedicated to the memory of 
	Professor Harald  Fritzsch}
\author{Tatsuo Kobayashi$^{\,1}$
		and~ Morimitsu Tanimoto$^{\,2}$
		\\*[20pt]
{
\begin{minipage}{\linewidth}
$^1${\it \normalsize Department of Physics, Hokkaido University, Sapporo 060-0810, Japan }
\\*[5pt]
$^2${\it \normalsize
Department of Physics, Niigata University, Ikarashi 2, 
Niigata 950-2181, Japan} 
\\*[5pt]
\end{minipage}
}
\\*[50pt]}
\date{
\centerline{\small \bf Abstract}
\begin{minipage}{0.9\linewidth}
\medskip
\medskip
\small
	We review the modular flavor symmetric models of quarks and leptons
focusing on  our works.
We present  some flavor models of quarks and leptons  
by using finite modular groups and discuss the phenomenological implications.
The modular flavor symmetry gives interesting phenomena at the fixed point of
modulus. As a representative, we show the successful texture structure at the fixed point $\tau = \omega$.
We also study CP violation,  which occurs  through the  modulus stabilization.
Finally, 
we study SMEFT with modular flavor symmetry by including higher dimensional operators.
\end{minipage}
}}
\begin{titlepage}
\maketitle
\thispagestyle{empty}
\end{titlepage}
\newpage
\section{Introduction}\label{ra_sec1}

One of important mysteries in particle physics is the origin of the flavor structure, 
i.e., fermion mass hierarchies, their mixing angles, and CP violation.
Various studies have been done to understand their origin.
One of traditional approaches is the texture zeros
proposed by Weinberg and Fritzsch, where zero entries are put in the fermion mass matrices
\cite{Weinberg:1977hb,Fritzsch:1977za}. 
Indeed, the Fritzsch Ansatz \cite{Fritzsch:1977vd,Fritzsch:1979zq} 
gave significant prediction power for the flavor mixing although the origin of zeros are unclear.
This approach leads to the texture zero analysis where some elements of mass matrices are required to be zero to reduce the degrees of freedom in mass matrices. 
Some famous works have been made in the texture zeros
\cite{Georgi:1979df,Branco:1988iq,Dimopoulos:1991za,Ramond:1993kv,Frampton:2002yf}.
Those zeros can be related with certain symmetries.

Flavor symmetries are interesting approaches to attack the origin of fermion mass hierarchies and 
mixing angles. Froggatt-Nielsen have taken  the $U(1)$ symmetry to  explain the observed masses and mixing angles for quarks \cite{Froggatt:1978nt}.
Furthermore, 
the  $S_3$ symmetry was used for quark mass matrices
\cite{Pakvasa:1977in,Wilczek:1977uh}.
It was also discussed  to understand the large mixing angle
\cite{Fukugita:1998vn} in the oscillation of atmospheric neutrinos \cite{Fukuda:1998mi}. 
For the last twenty years, non-Abelian discrete flavor symmetries  have been developed, that is
motivated by the precise observation of  flavor mixing angles of  leptons
\cite{
	Altarelli:2010gt,Ishimori:2010au,Ishimori:2012zz,Hernandez:2012ra,
	King:2013eh,King:2014nza,Tanimoto:2015nfa,King:2017guk,Petcov:2017ggy,Kobayashi:2022moq,Feruglio:2019ktm}.

The standard model (SM) is the low-energy effective field theory from the viewpoint of 
underlying theory, and it is referred to as the SM effective field theory (SMEFT) \cite{Buchmuller:1985jz,Grzadkowski:2010es,Alonso:2013hga}.
The SMEFT includes many higher dimensional operators, and they contribute to flavor changing processes and 
muon $(g-2)$.
Flavor symmetries are useful not only to derive realistic fermion masses and their mixing angles, but 
to control higher dimensional operators in the SMEFT.
Indeed, the $U(3)^5$ and  $U(2)^5$ symmetries control the SMEFT operators
\cite{Faroughy:2020ina}.
The $U(3)^5$ symmetry \cite{Gerard:1982mm} allows us to apply the Minimal Flavor Violation (MFV) hypothesis~\cite{Chivukula:1987py,DAmbrosio:2002vsn}, 
which is the most restrictive hypothesis consistent with the SMEFT. 
In the $U(2)^5$ symmetry~\cite{Barbieri:2011ci,Barbieri:2012uh,Blankenburg:2012nx}, 
it retains most of the MFV virtues and allows us to have a much richer structure  as far as the dynamics of third family is concerned. 
Thus,
flavor symmetries are useful to connect between the low-energy physics and 
high-energy physics such as superstring theory.

Superstring theory is a promising candidate for unified theory of all the interactions including gravity 
and matters such as quarks and leptons, and Higgs modes.
Superstring theory must have six-dimensional (6D) compact space in addition to 
four-dimensional (4D) space times.
Geometrical symmetries of 6D compact space control 4D effective field theory.
For example, in certain compactifications there appears non-Abelian discrete symmetries such as 
$D_4$ and $\Delta(54)$ \cite{Kobayashi:2004ya,Kobayashi:2006wq,Ko:2007dz,Abe:2009vi,Beye:2014nxa}.


The modular symmetry is a geometrical symmetry of $T^2$ and $T^2/Z_2$, and corresponds to  change of their cycle basis.
Matter modes transform non-trivially under the modular symmetry.
(See for hetetrotic string theory on orbifolds Refs.~\cite{Ferrara:1989qb,Lerche:1989cs,Lauer:1989ax,Lauer:1990tm} 
and magnetized D-brane models Refs.~\cite{Kobayashi:2018rad,Kobayashi:2018bff,Ohki:2020bpo,Kikuchi:2020frp,Kikuchi:2020nxn,
	Kikuchi:2021ogn,Almumin:2021fbk}\,.\ \footnote{Calai-Yau manifolds have larger symplectic 	modular symmetries of many moduli \cite{Strominger:1990pd,Candelas:1990pi,Ishiguro:2020nuf,Ishiguro:2021ccl}\,.})
That is, the modular symmetry is a flavor symmetry.
Indeed, finite modular groups include $S_3$, $A_4$, $S_4$, $A_5$, which have been used in 4D flavor models so far, while 
$\Delta(98)$ and $\Delta(384)$ are also included as subgroups.

The well-known finite groups $S_3$, $A_4$, $S_4$ and $A_5$
are isomorphic to the finite modular groups 
$\Gamma_N$ for $N=2,3,4,5$, respectively\cite{deAdelhartToorop:2011re}.
The lepton mass matrices have been presented in terms of {$\Gamma_3\simeq A_4$} modular forms \cite{Feruglio:2017spp}.
Modular forms have also been obtained for $\Gamma_2\simeq  S_3$ \cite{Kobayashi:2018vbk},
$\Gamma_4 \simeq  S_4$ \cite{Penedo:2018nmg} and  
$\Gamma_5 \simeq  A_5$ \cite{Novichkov:2018nkm,Ding:2019xna}, respectively.
By using them, the viable lepton mass matrices have been obtained
for $\Gamma_4 \simeq  S_4$ \cite{Penedo:2018nmg}, and then
$\Gamma_5 \simeq  A_5$ \cite{Novichkov:2018nkm,Ding:2019xna}.

The 4D CP symmetry can be embedded into a proper Lorentz symmetry in higher dimensional theory such as 
superstring theory \cite{Green:1987mn,Strominger:1985it,Dine:1992ya,Choi:1992xp,Lim:1990bp,Kobayashi:1994ks}.
From this viewpoint, CP violation in 4D effective field theory would originate from the compactification, that is, 
the moduli stabilization. (See for early studies on the CP violation through the moduli stabilization Refs.~\cite{Acharya:1995ag,Dent:2001cc,Khalil:2001dr,Giedt:2002ns}\,.)
Recently, the spontaneous breaking of the CP symmetry was studied 
through the moduli stabilization due to 3-form fluxes Refs.~\cite{Kobayashi:2020uaj,Ishiguro:2020nuf}.
In modular flavor symmetric models, the CP symmetry is combined with the modular symmetry as well as other symmetries, and 
is enlarged \cite{Baur:2019kwi,Novichkov:2019sqv,Baur:2019iai,Baur:2020jwc,Nilles:2020gvu,Ishiguro:2021ccl}\ \footnote{See for 
	the CP symmetry in the Calabi-Yau compactification Refs.~\cite{Ishiguro:2020nuf,Ishiguro:2021ccl,Bonisch:2022slo}\,.}.
The CP-invariant vacua and CP-preserving modulus values increase by the modular symmetry.
It is important to study the CP violation in such models with the enlarged symmetry \cite{Kobayashi:2019uyt,Kikuchi:2022geu}.

Higher dimensional operators can be computed within the framework of superstring theory.
Allowed couplings are controlled by stringy symmetries and $n$-point couplings are written by products of 3-point couplings.
The modular flavor symmetry also control these higher dimensional operators \cite{Kobayashi:2021uam,Kobayashi:2021pav,Kobayashi:2022jvy}.

In addition to the above aspects, 
the modular flavor symmetries were recently extended to models for dark matter \cite{Nomura:2019jxj,Kobayashi:2021ajl}, 
soft supersymmetry breaking terms \cite{Kobayashi:2021jqu,Tanimoto:2021ehw,Kikuchi:2022pkd}, matter parity 
\cite{Kobayashi:2022sov}, the strong CP problem\cite{Feruglio:2023uof}, etc.

The paper is organized as follows.
In section 2, we give a brief review on modular symmetry.
In section 3, we study modular flavor symmetric models.
As an illustrating example, we explain $A_4$ modular symmetric models.
In section 4, we study texture structure at the fixed point $\tau = \omega$.
In section 5, we study CP violation in modular symmetric models.
In section 6, we study SMEFT with modular flavor symmetry.
Section 7 is devoted to conclusion.
In Appendix \ref{Modularforms}, we review modular forms of $A_4$.
In Appendix \ref{Tensor},  the tensor product
decomposition is given in the  $A_4$ group.

\section{Modular Symmetry}
\begin{figure}[h]
	\begin{center}
		\includegraphics[width=0.5\linewidth]{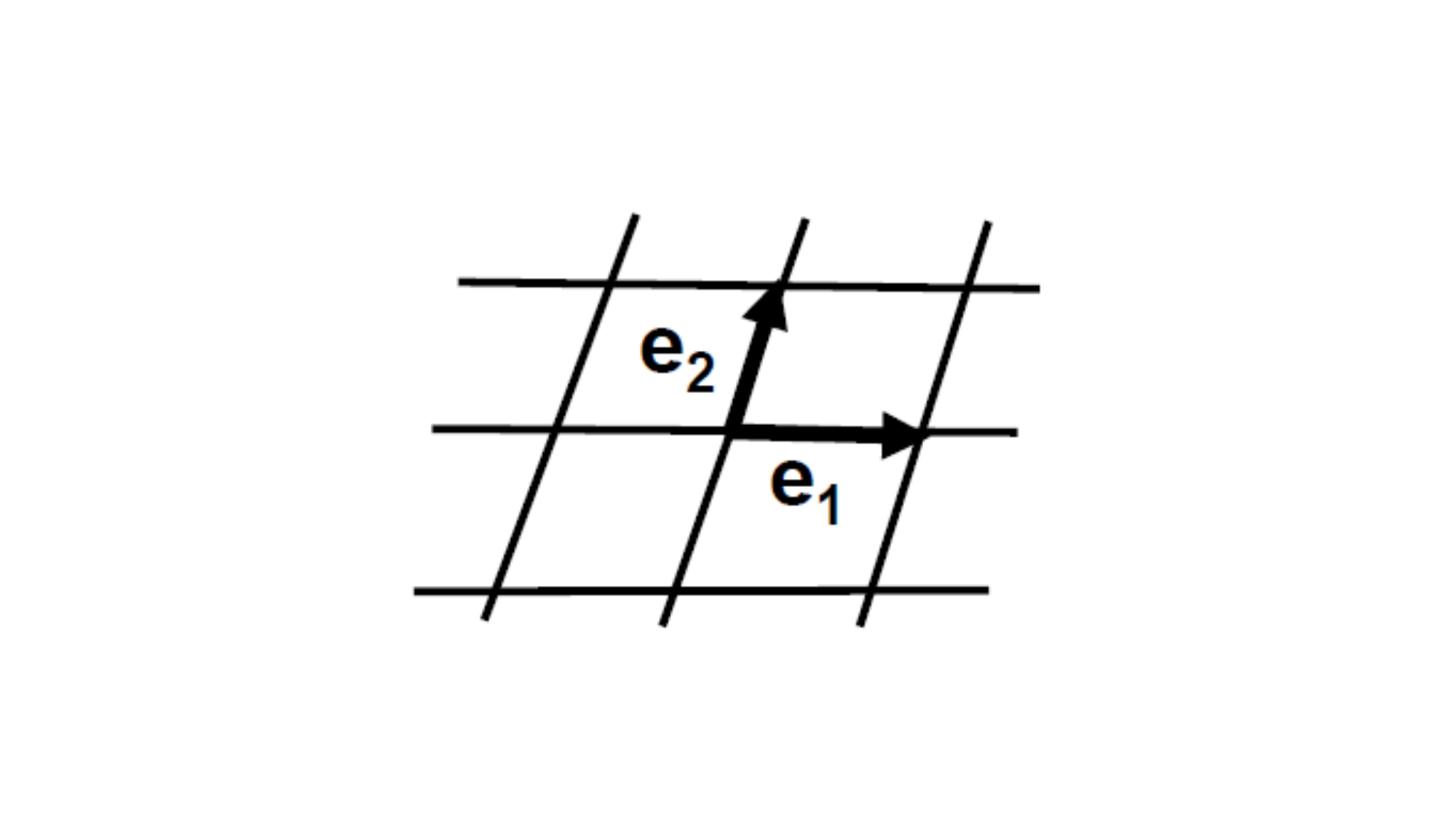}
	\end{center}
	\vskip -1 cm
	\caption{Lattice $\Lambda$ and two basis vectors, $e_1$ and $e_2$.}
	\label{fig:Lattice}
\end{figure}

The {\it modular symmetry} is a geometrical symmetry of the two-dimensional torus, $T^2$.
The two-dimensional torus is constructed as division of the two-dimensional Euclidean space $R^2$ by 
a lattice $\Lambda$, $T^2=R^2/\Lambda$.
Instead of $R^2$, one can use the one-dimensional complex plane.
As shown in Fig.\,\ref{fig:Lattice}, the lattice is spanned by two basis vectors, $e_1$ and $e_2$ as $m_1e_1+m_2e_2$, where $m_1$ and $m_2$ are integer.
Their ratio,
\begin{eqnarray}
\tau =\frac{e_2}{e_1},
\label{eq:tau}
\end{eqnarray}
in the complex plane, represents the shape of $T^2$, and the parameter $\tau$ is 
called the {\it modulus}.


The same lattice can be spanned by other basis vectors such as
\begin{eqnarray}
\left(
\begin{array}{c}
e'_2 \\ e'_1 
\end{array}\right)
=\left(
\begin{array}{cc}
a &  b \\ c & d
\end{array}\right)
\left(
\begin{array}{c}
e_2 \\ e_1 
\end{array}\right),
\label{eq:lattice-basis}
\end{eqnarray}
where $a,b,c,d$ are integer satisfying $ad-bc=1$.
That is the $SL(2,Z)$.

Under the above transformation, 
the  modulus $\tau$ transforms as follows,
\begin{eqnarray}
\tau \longrightarrow \tau'=\gamma \tau = \frac{a\tau +b}{c\tau +d}\,.
\label{eq:tau-transformation}
\end{eqnarray}
That is the modular symmetry \cite{Gunning:1962,Schoeneberg:1974,Koblitz:1984,Bruinier:2008}.
For the element $-e$ in $SL(2,Z)$,
\begin{eqnarray}
-e=\left(
\begin{array}{cc}
-1 & 0 \\ 0 & -1
\end{array}\right),
\label{eq:-1}
\end{eqnarray}
the modulus $\tau$ is invariant, 
$\tau \longrightarrow \tau'=(-\tau)/(-1)=\tau$.
Thus, the modular group is $\bar \Gamma=PSL(2,Z)=SL(2,Z)/\{e,-e \}$.
It is sometimes called the {\it inhomogeneous modular group}.
On the other hand, the group, $\Gamma=SL(2,Z)$ is called the {\it homogeneous modular group} or 
the {\it full modular group}.

\begin{figure}[h]
	\begin{minipage}[]{0.47\linewidth}
		\begin{center}
			\includegraphics[width=0.8\linewidth]{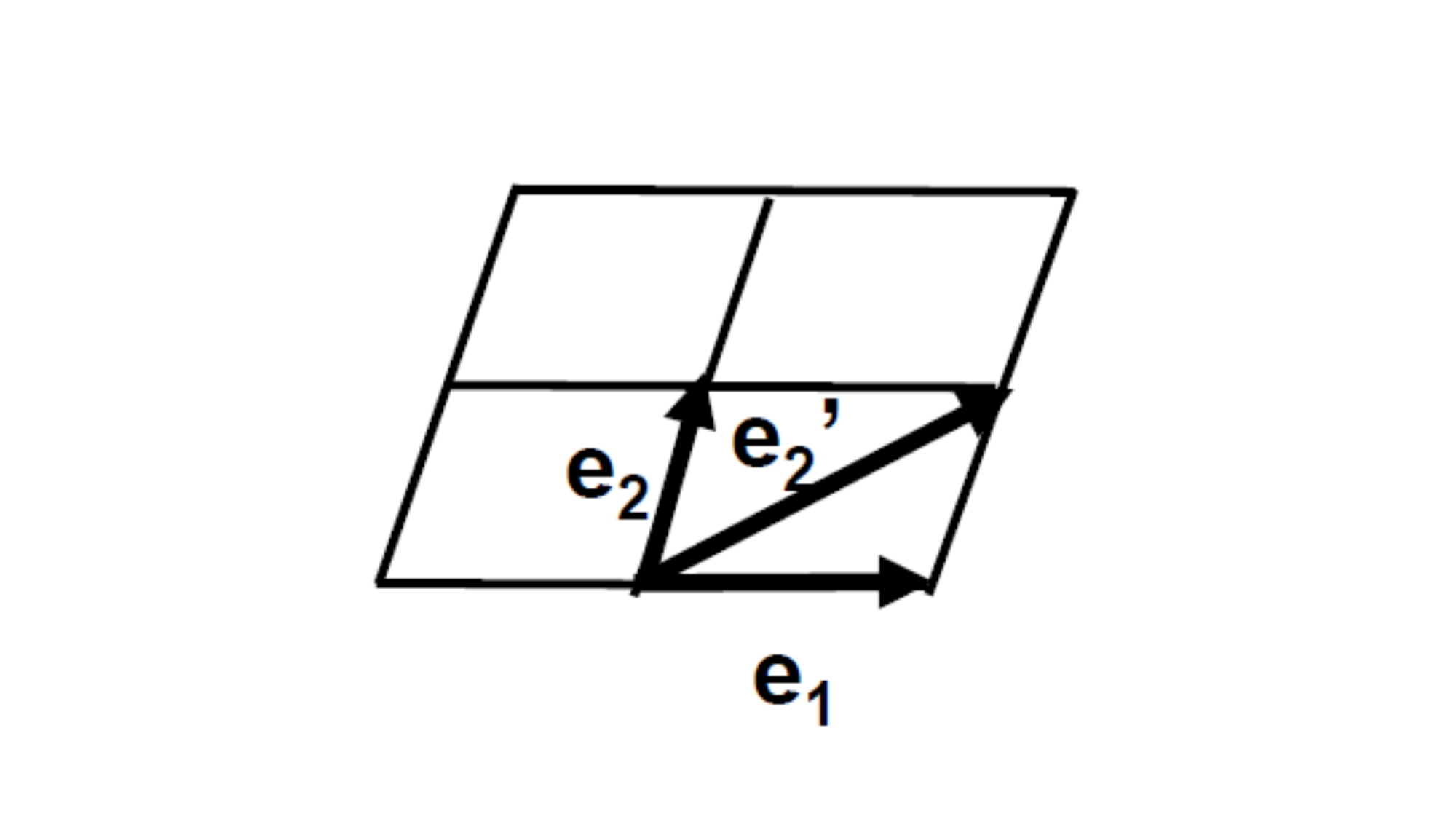}
		\end{center}
		\vskip -0.5 cm
		\caption{Basis change with $(a,b,c,d)=(1,1,0,1)$ corresponds to $T$.}
		\label{fig:T}
	\end{minipage}
	\hspace{5mm}
	\begin{minipage}[]{0.47\linewidth}
	\begin{center}
			\includegraphics[width=0.8\linewidth]{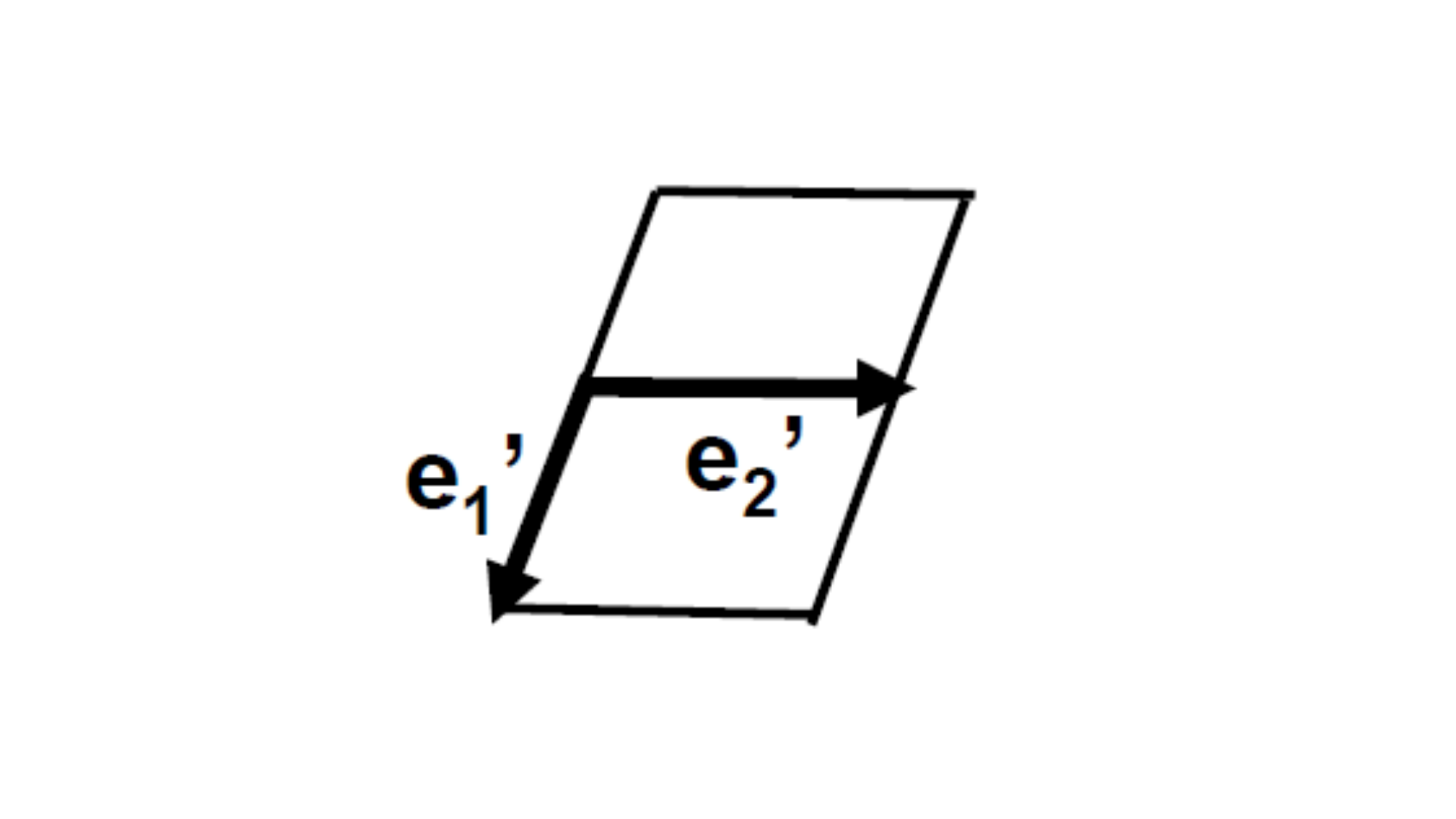}
		\end{center}
		\vskip -0.5 cm
		\caption{Basis change with $(a,b,c,d)=(0,1,-1,0)$ corresponds to $S$.}
		\label{fig:S}
	\end{minipage}
\end{figure}

The generators of $\Gamma \simeq SL(2,Z)$ are written by $S$ and $T$,
\begin{eqnarray}
S=\left(
\begin{array}{cc}
0 & 1 \\ -1 & 0
\end{array}
\right), 
\qquad 
T=\left(
\begin{array}{cc}
1 & 1 \\ 0 & 1
\end{array}
\right).
\label{eq:S-T}
\end{eqnarray}
They satisfy the following algebraic relations,
\begin{eqnarray}
S^4=(ST)^3=e.
\label{eq:ST-SL}
\end{eqnarray}
Note that 
\begin{eqnarray}
S^2=-e.
\label{eq:S^2}
\end{eqnarray}
On $\bar \Gamma=PSL(2,Z)$, they satisfy 
\begin{eqnarray}
S^2=(ST)^3=e.
\label{eq:ST-PSL}
\end{eqnarray}
These relations are also confirmed explicitly by 
the following transformations:
\begin{eqnarray}
S:~\tau \longrightarrow -\frac{1}{\tau},\qquad T:~\tau \longrightarrow \tau +1 ,
\label{eq:ST-tau}
\end{eqnarray}
which  are shown on  the lattice $\Lambda$ in Figs.\,\ref{fig:T} and \ref{fig:S}.


In addition to the above algebraic relations of $\bar \Gamma=PSL(2,Z)$, 
we can require $T^N=e$, i.e.
\begin{eqnarray}
S^2=(ST)^3=T^N=e.
\label{eq:T^N}
\end{eqnarray}
They can correspond to finite groups such as 
$S_3, A_4, S_4, A_5$ for $N=2,3,4,5$.
In practice, we define the {\it principal congruence subgroup} $\Gamma(N)$ as 
\begin{eqnarray}
\Gamma(N) =\left\{
\left(
\begin{array}{cc}
a &  b \\ c & d
\end{array}\right) \in {\Gamma}~ \right| \left. 
\left(
\begin{array}{cc}
a &  b \\ c & d
\end{array}\right) = 
\left(
\begin{array}{cc}
1 &  0 \\ 0 & 1
\end{array}\right)~~({\rm mod}~N) \right\}.
\label{eq:Gamma(N)}
\end{eqnarray}
It includes $T^N$, but not $S$ or $T$.
Then, we define the quotient $\Gamma_N=\bar \Gamma/\bar \Gamma(N)$, where 
the above algebraic relations are satisfied.
It is found that $\Gamma_N$ with $N=2,3,4,5$ are isomorphic to 
$S_3, A_4, S_4, A_5$, respectively \cite{deAdelhartToorop:2011re}.

We define $SL(2,Z_N)$ by 
\begin{eqnarray}
SL(2,Z_N) = \left. \left\{
\left(
\begin{array}{cc}
a &  b \\ c & d
\end{array}\right)  \right| 
a,b,c,d \in Z_N, ad-bc=1 \right\},
\label{eq:SL(2,N)}
\end{eqnarray}
where $Z_N$ denotes integers modulo $N$.
The group $\Gamma_N$ is isomorphic to $PSL(2,Z_N) = SL(2,Z_N)/\{ e,-e \}$ for $N>2$, 
while $\Gamma_2$ is isomorphic to $SL(2,Z_2)$, because $e=-e$ in $SL(2,Z_2)$.

Similar to $\Gamma_N$, we can define $\Gamma'_N=SL(2,Z)/\Gamma(N)$, 
and it is the double cover of $\Gamma_N$.
That is, the groups $\Gamma'_N$ for $N=3,4,5$ are isomorphic to the double covering groups of 
$A_4,S_4,A_5$, i.e. $T',S'_4, A'_5$, respectively, 
although $\Gamma_2'$ is isomorphic to $S_3$.

The upper half-plane of the modulus space $\tau$ is mapped onto itself.
For example,  $\Gamma$ does not include the basis change.  $(e_1,e_2) \longrightarrow (e_1,-e_2)$, 
i.e. $\tau \to -\tau$.
In practice, we find 
\begin{eqnarray}
{\rm Im}(\gamma \tau) = |c\tau +d|^{-2}{\rm Im}(\tau).
\label{eq:upper-half}
\end{eqnarray}
Thus, the modular group is represented on the upper half-plane of $\tau$.
Obviously one can map any value of $\tau$ on the upper half-plane into the region, $-\frac12 \leq {\rm Re}(\tau) \leq \frac12 $ 
by $T^n$.
Furthermore, by the modular transformation one can map any value of $\tau$ on the upper half-plane into the following region:
\begin{eqnarray}
-\frac12 \leq {\rm Re}(\tau) \leq \frac12, \qquad |\tau|>1,
\label{eq:fundamental}
\end{eqnarray}
which is called the {\it fundamental domain}.
Suppose that ${\rm Im}(\gamma \tau)$ is a maximum value among all of $\gamma$ for a fixed value of $\tau$.
If $|\gamma \tau|<1$, we map it by $S$, and we find 
\begin{eqnarray}
{\rm Im}(S\gamma \tau)=\frac{{\rm Im}(\gamma \tau)}{|\gamma \tau|^2}> {\rm Im}(\gamma \tau).
\label{eq:S-gamma}
\end{eqnarray}
That is inconsistent with the assumption that ${\rm Im}(\gamma \tau)$ is a maximum value among all of $\gamma$.
That is,  we find  $|\gamma \tau| >1$.
Thus, we can map $\tau$ on the upper half-plane into the fundamental region by the modular transformation.
The point $\tau =i$ is the fixed point under $S$ because $S:i\to -\frac 1i=i$, where $Z_2$ symmetry remains.
Similarly, the point $\tau=e^{2\pi i/3}$ is the fixed point under $ST$, where $Z_3$ symmetry remains.

Modular forms $f_i(\tau)$ of weight $k$ are the holomorphic functions of $\tau$ and transform as
\begin{equation}
f_i(\tau) \longrightarrow (c\tau +d)^k \rho(\gamma)_{ij}f_j( \tau)\, ,
\quad \gamma\in \bar \Gamma\, ,
\label{modularforms}
\end{equation}
under the modular symmetry, where
$\rho(\gamma)_{ij}$ is a unitary matrix under $\Gamma_N$.

Under the modular transformation, chiral superfields $\psi_i$ ($i$ denotes flavors) with weight $-k$
transform as \cite{Ferrara:1989bc}
\begin{equation}
\psi_i\longrightarrow (c\tau +d)^{-k_i}\rho(\gamma)_{ij}\psi_j\, .
\label{chiralfields}
\end{equation}

We study global SUSY models.
The superpotential which is built from matter fields and modular forms
is assumed to be modular invariant, i.e., to have 
a vanishing modular weight. For given modular forms, 
this can be achieved by assigning appropriate
weights to the matter superfields.

The kinetic terms  are  derived from a K\"ahler potential.
The K\"ahler potential of chiral matter fields $\psi_i$ with the modular weight $-k$ is given simply  by 
\begin{equation}\label{eq:Kahler}
\frac{1}{[i(\bar\tau - \tau)]^{k}} \sum_i|\psi_i|^2,
\end{equation}
where the superfield and its scalar component are denoted by the same letter, and $\bar\tau =\tau^*$ after taking vacuum expectation value (VEV) of $\tau$.
The canonical form of the kinetic terms  is obtained by 
changing the normalization of parameters.
The general K\"ahler potential consistent with the modular symmetry possibly contains additional terms \cite{Chen:2019ewa}. However, we consider only the simplest form of
the K\"ahler potential.

\section{Modular flavor symmetric models}

In this section, we discuss the  flavor model of quark and lepton mass matrices.
There is a difference between the modular symmetry and the usual flavor symmetry.
Coupling constants such as Yukawa couplings also transform non-trivially
under the modular symmetry and are written as functions of the modulus called modular forms, which are   holomorphic functions of  the modulus  $\tau$.
On the other hand,  coupling constants are invariant under the traditional flavor symmetries.

The  flavor model of lepton mass matrices have been proposed
based on  the finite modular group  $\Gamma_3 \simeq A_4$ \cite{Feruglio:2017spp}.
This   approach based on modular invariance
opened up a new promising direction in the studies of the flavor physics and correspondingly in flavor model building.

\subsection{Modular $A_4$ invariance and neutrino mixing}
We present a phenomenological discussion of the modular invariant lepton mass matrix
by using  the finite modular group  $\Gamma_3 \simeq A_4$,
where a simple model was proposed by Feruglio \cite{Feruglio:2017spp}.
We have shown that it can predict a clear correlation between the neutrino mixing angle $\theta_{23}$ and the CP violating Dirac phase \cite{Kobayashi:2018scp}.

The mass matrices of neutrinos and charged leptons
are essentially given  by fixing the expectation value of the  modulus $\tau$,
which is the only source of  the  breaking of the modular invariance.
Since there are freedoms for the assignment of irreducible representations and modular weights to leptons,
suppose that 
three left-handed lepton doublets are of a triplet of the $A_4$ group.
The three right-handed neutrinos are also of a triplet of $A_4$.
On the other hand, the Higgs doublets are supposed to be a trivial singlet of $A_4$ for simplicity (In the next section, we modify this assumption.).
We also  assign three right-handed charged leptons for three different singlets of $A_4$, $(1,1',1'')$,  respectively.
Therefore, there are three independent couplings in the superpotential of the charged lepton sector.
Those coupling constants can be adjusted to the  observed charged lepton masses.

The assignments of representations and modular weights to the MSSM fields as well as right-handed neutrino superfields 
are presented in Table \ref{tb:fields}.

\begin{table}[h]
	\centering
	\begin{tabular}{|c|c|c|c|c|c|c|} \hline 
		\rule[12pt]{0pt}{0pt}
		&$L$&$e^c,\mu^c,\tau^c$&$\nu^c$&$H_u$&$H_d$&${\bf Y_3}=(Y_1,Y_2,Y_3)^T$\\ \hline 
		\rule[14pt]{0pt}{0pt}
		SU(2)&$2$&$1$&$1$&$2$&$2$&$1$\\
		$A_4$&$3$& 1,\ 1'',\ 1'&$3$&$1$&$1$&$3$\\
		$k_I$&$1$&$1$&$1$&0&0&$k=2$ \\ \hline
	\end{tabular}
	\caption{
		The charge assignment of SU(2), $A_4$, and the modular weight. 	}
	\label{tb:fields}
\end{table}

In terms of modular forms of $A_4$ triplet, ${\bf Y_3}$ 
in  Eq.\eqref{weight2} of Appendix \ref{Modularforms},
the modular invariant Yukawa coupling and Majorana mass terms of the leptons are given by the following superpotentials:
\begin{eqnarray}
w_e&&=\alpha_e H_d(L{\bf Y_3})e^c+\beta_e H_d(L{\bf Y_3})\mu^c+\gamma_e H_d(L{\bf Y_3})\tau^c~,\label{charged} \\
w_D&&=g_i( H_u L \,\nu^c\,{\bf Y_3})_{\bf 1}~,  \label{Dirac}\\
w_N&&=\Lambda(\nu^c\nu^c{\bf Y_3})_{\bf 1}~, \label{Majorana}
\end{eqnarray}
where the sums of the modular weights  vanish.
The parameters $\alpha_e$,  $\beta_e$,  $\gamma_e$,  $g_i$($i=1,2$), and $\Lambda$
are constant coefficients.

VEVs of the neutral component of $H_u$ and $H_d$
are written  as $v_u$ and $v_d$, respectively.
Then, the mass matrix of charged leptons is given by
the superpotential  Eq.\,(\ref{charged}) as follows:
\begin{align}
\begin{aligned}
M_E&=v_d\,
\begin{pmatrix}
Y_1 & Y_2 & Y_3 \\
Y_3 & Y_1 & Y_2 \\
Y_2 & Y_3 & Y_1
\end{pmatrix}
\begin{pmatrix}
\alpha_e & 0 &0  \\
0&\beta_e&0\\
0 & 0& \gamma_e
\end{pmatrix}_{LR}\,.
\end{aligned}\label{eq:CL}
\end{align}
The coefficients $\alpha_e$, $\beta_e$, and $\gamma_e$ 
are taken to be real positive by rephasing  right-handed charged lepton fields
without loss of generality.


Since the tensor product of $3\otimes 3$ is decomposed  into the symmetric triplet and the antisymmetric triplet as seen in Appendix \ref{Tensor},  the superpotential of
the Dirac neutrino mass  in Eq.\,(\ref{Dirac}) is expressed by introducing additional two parameters
$g_1$ and $g_2$ as:
\begin{align}
\begin{aligned}
w_D=&v_u\left[
g_1\begin{pmatrix}
2\nu_eY_1-\nu_\mu Y_3-\nu_\tau Y_2\\
2\nu_\tau Y_3-\nu_e Y_2-\nu_\mu Y_1\\
2\nu_\mu Y_2-\nu_\tau Y_1-\nu_eY_3\end{pmatrix} \oplus
g_2\begin{pmatrix}\nu_\mu Y_3-\nu_\tau Y_2\\\nu_e Y_2-\nu_\mu Y_1\\\nu_\tau Y_1-\nu_e Y_3\end{pmatrix}\right]  \otimes
\begin{pmatrix}\nu^c_{1}\\\nu^c_{2}\\\nu^c_{3}\end{pmatrix}\\
=&v_ug_1\left[
(2\nu_eY_1-\nu_\mu Y_3-\nu_\tau Y_2)\nu^c_{1}+
(2\nu_\mu Y_2-\nu_\tau Y_1-\nu_eY_3)\nu^c_{2} \right .\\
&\left .+(2\nu_\tau Y_3-\nu_e Y_2-\nu_\mu Y_1)\nu^c_{3}\right] \\
&+v_ug_2\left[
(\nu_\mu Y_3-\nu_\tau Y_2)\nu^c_{1}+
(\nu_\tau Y_1-\nu_e Y_3)\nu^c_{2}+
(\nu_e Y_2-\nu_\mu Y_1)\nu^c_{3}\right].
\end{aligned}
\end{align}
The  Dirac neutrino mass matrix is given as
\begin{align}
M_D=v_u\begin{pmatrix}
2g_1Y_1 & -(g_1-g_2)Y_3 & \ - (g_1+g_2)Y_2 \\
-(g_1+g_2)Y_3 & 2g_1Y_2 &\ -(g_1-g_2)Y_1 \\
-(g_1-g_2)Y_2 &\ -(g_1+g_2)Y_1 & 2g_1Y_3\end{pmatrix}_{LR}.
\label{MD}
\end{align}


On the other hand,
since the Majorana neutrino mass terms are symmetric,
the superpotential in  Eq.\,(\ref{Majorana}) is expressed simply as
\begin{align}
\begin{aligned}
w_N=&\Lambda\begin{pmatrix}
2\nu^c_{1}\nu^c_{1}-\nu^c_{2}\nu^c_{3}-\nu^c_{3}\nu^c_{2}\\
2\nu^c_{3}\nu^c_{3}-\nu^c_{1}\nu^c_{2}-\nu^c_{2}\nu^c_{1}\\
2\nu^c_{2}\nu^c_{2}-\nu^c_{3}\nu^c_{1}-\nu^c_{1}\nu^c_{3}\end{pmatrix}\otimes
\begin{pmatrix}Y_{1}\\Y_{2}\\Y_{3}\end{pmatrix} \\
=&\Lambda\left[(2\nu^c_{1}\nu^c_{1}-\nu^c_{2}\nu^c_{3}-
\nu^c_{3}\nu^c_{2})Y_1+
(2\nu^c_{3}\nu^c_{3}-\nu^c_{1}\nu^c_{2}-\nu^c_{2}\nu^c_{1})Y_3\right. \\
&\left.+(2\nu^c_{2}\nu^c_{2}-\nu^c_{3}\nu^c_{1}-\nu^c_{1}\nu^c_{3})Y_2\right].
\end{aligned}
\end{align}
Then, the modular invariant right-handed Majorana neutrino mass matrix is given as
\begin{align}
M_N=\Lambda\begin{pmatrix}
2Y_1 & -Y_3 & -Y_2 \\
-Y_3 & 2Y_2 & -Y_1 \\
-Y_2 & -Y_1 & 2Y_3\end{pmatrix}_{RR}.
\end{align}
Finally, the effective neutrino mass matrix is obtained by the type I seesaw 
as follows:
\begin{align}
M_\nu=-M_D^{}M_N^{-1}M_D^{\rm T} ~.
\end{align}


When $\tau$ is fixed, the modular invariance is broken, and then the lepton mass matrices give the mass eigenvalues and flavor mixing numerically.
In order to determine the value of  $\tau$, we use the result of NuFIT 5.1 \cite{Esteban:2020cvm}.
By inputting the data of $\Delta m_{\rm atm}^2 \equiv m_3^2-m_1^2$, $\Delta m_{\rm sol}^2 \equiv  m_2^2-m_1^2$, and 
three mixing angles $\theta_{23}$, $\theta_{12}$, and $\theta_{13}$ with $3\,\sigma$ error-bar,   
we can severely constraint values of  the modulus $\tau$ and the other parameters, and then we can predict the CP violating Dirac phases $\delta_{CP}$.
We consider both the normal hierarchy (NH) of neutrino masses $m_1<m_2<m_3$ 
and the inverted hierarchy (IH) of neutrino masses $m_3<m_1<m_2$, where
$m_1$, $m_2$, and $m_3$ denote three light neutrino masses.
Since the sum of three neutrino  masses  $\sum m_i$  is constrained
by the recent cosmological data \cite{Vagnozzi:2017ovm,Aghanim:2018eyx,ParticleDataGroup:2022pth},
we exclude the predictions with $\sum m_i\geq 200$\,meV 
even if mixing angles are consistent with observed one.


The coefficients $\alpha_e/\gamma_e$ and  $\beta_e/\gamma_e$ in the charged lepton mass matrix are given only in terms of $\tau$ by the input of the observed values  $m_e/m_\tau$ and $m_\mu/m_\tau$.
As the input charged lepton masses, 
we take Yukawa couplings of charged leptons 
at the GUT scale $2\times 10^{16}$ GeV,  where $\tan\beta=5$ is taken
as a bench mark
\cite{Antusch:2013jca, Bjorkeroth:2015ora}:
\begin{eqnarray}
&&y_e=(1.97\pm 0.024) \times 10^{-6}, \quad 
y_\mu=(4.16\pm 0.050) \times 10^{-4}, \nonumber\\
&&y_\tau=(7.07\pm 0.073) \times 10^{-3}.
\label{leptonyukawa}
\end{eqnarray}
Lepton masses are  given by $m_\ell=y_\ell v_H$ with $v_H=174$ GeV.

Then,
we have two free complex  parameters,  $g_2/g_1$ and the modulus $\tau$  apart from the overall factors in the neutrino sector.
The value of $\tau$ is scanned in the fundamental domain of $SL(2,Z)$. 

\begin{figure}[h!]
	\begin{tabular}{ccc}
		\begin{minipage}{0.47\hsize}
			\includegraphics[width=1.1\linewidth]{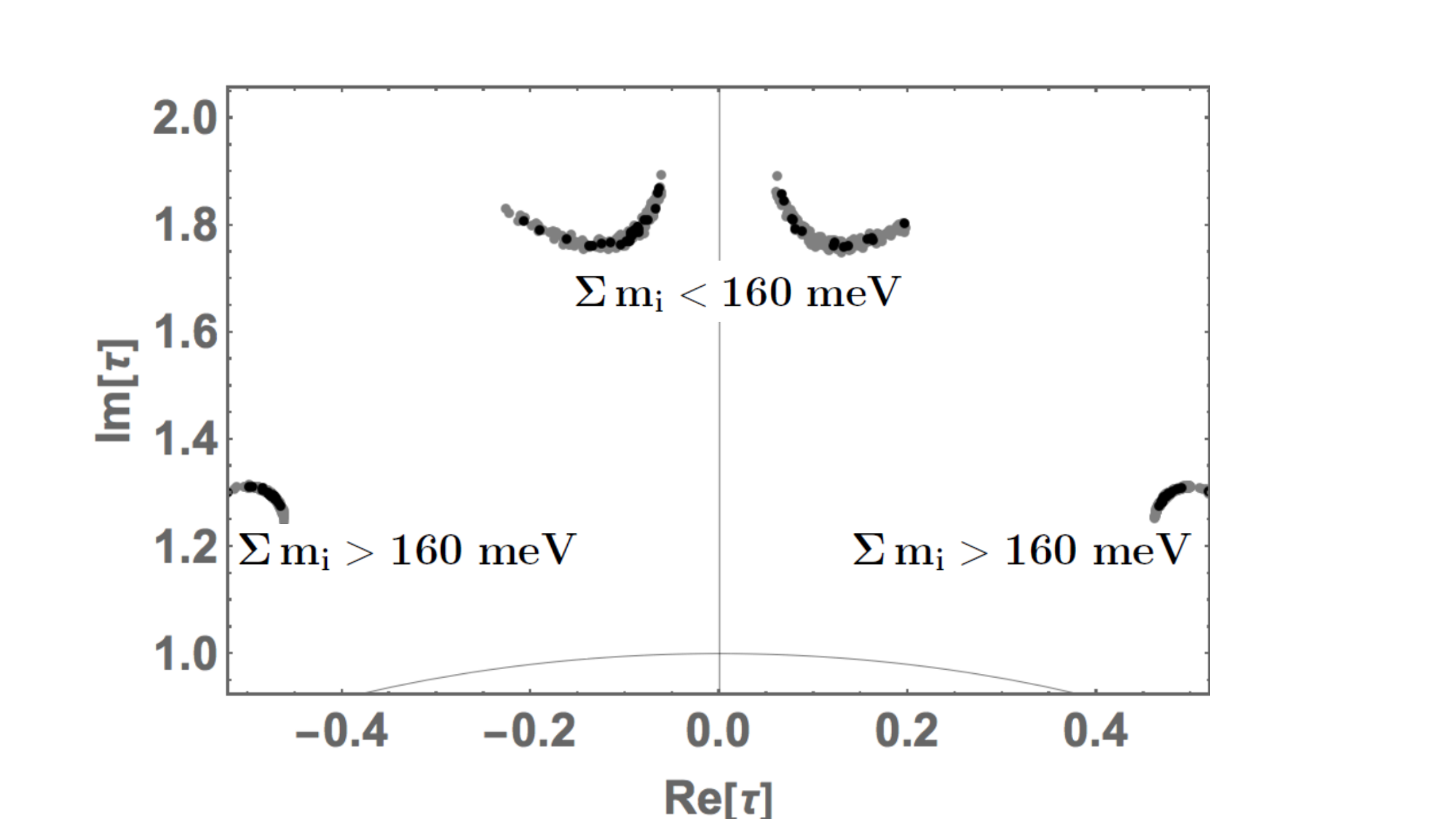}
			\caption{The allowed regions of ${\rm Re}[\tau]$ and ${\rm Im}[\tau]$  for NH.
				The predicted regions 
				$\sum m_i< 160$ meV and $\sum m_i> 160$ meV
				are shown, respectively.
				Black and grey points are in $2\,\sigma$ and $3\,\sigma$ regions,
				respectively.}
			\label{tauplot}
		\end{minipage}
		\phantom{=}
		\begin{minipage}{0.47\linewidth}
			\includegraphics[width=1.1\linewidth]{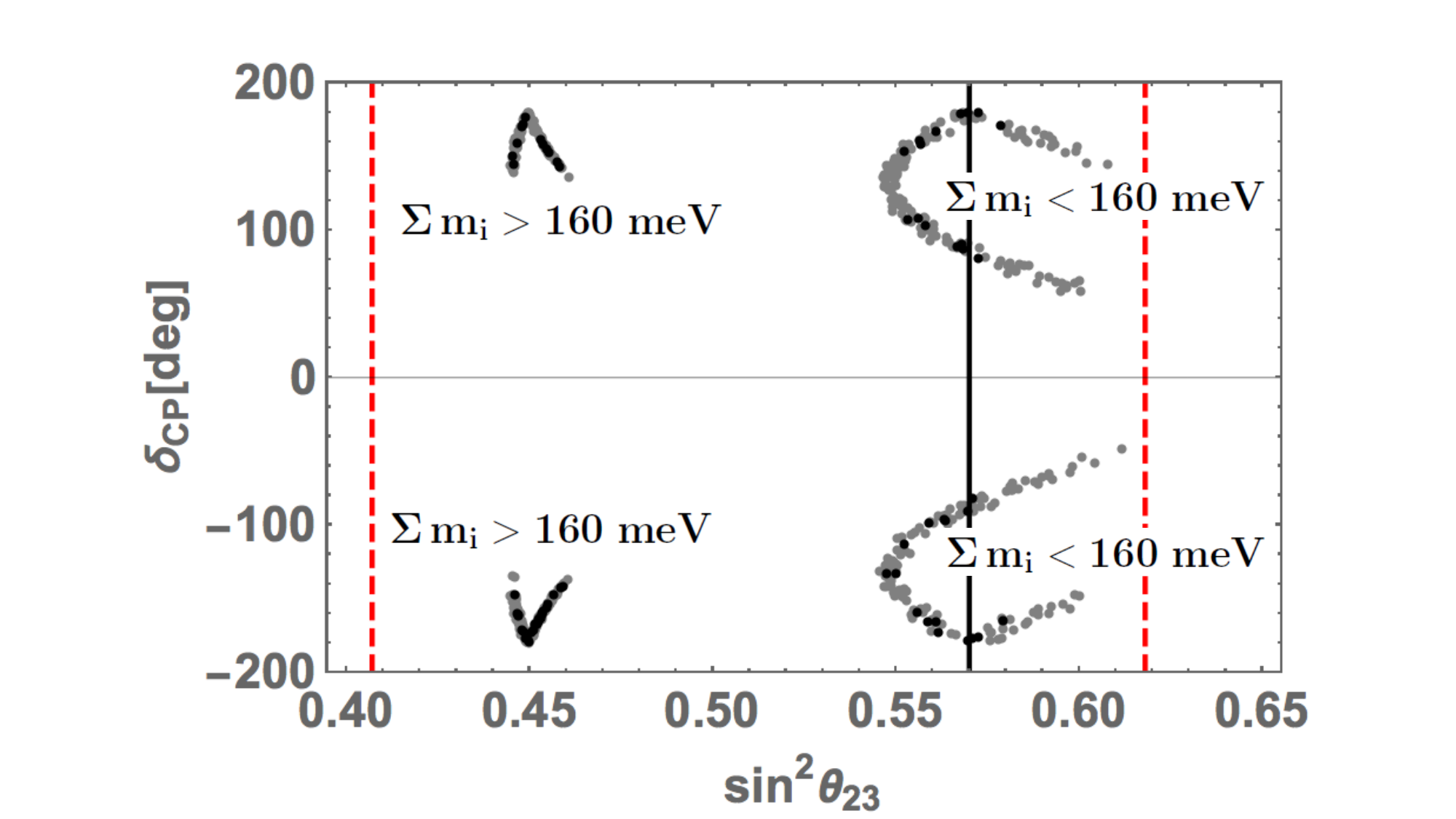}
			\caption{Predictions of $\delta_{CP}$ versus $\sin^2\theta_{23}$ for NH. 	Vertical solid and dashed lines  denote the central value,
				and  the upper and lower bounds of the experimental data with 
				$3 \ \sigma$, respectively.
				Notations are same as in Fig.\ref{tauplot}.}
			\label{Diracphase}
		\end{minipage}
	\end{tabular}
\end{figure}

In Fig.\ref{tauplot},
we show  allowed regions  
in the ${\rm Re}[\tau]$-${\rm Im}[\tau]$ plain for  NH of neutrino masses.
Those are given in the sum of neutrino masses
$\sum m_i=140$--$150$\,meV 
and $\sum m_i=170$--$200$\,meV.
The $2\,\sigma$ and $3\,\sigma$ regions are presented by
black and grey points, respectively.
If the cosmological observation confirms $\sum m_i< 160$ meV,
the region of  $\tau$ is severely restricted in this model.

We present the prediction of the Dirac CP violating phase $\delta_{CP}$ versus $\sin^2\theta_{23}$  for NH of neutrino masses in Fig.\ref{Diracphase}.
The predicted regions  correspond to regions of $\tau$ in Fig.\ref{tauplot}.
It is emphasized that 
the  predicted  $\sin^2\theta_{23}$ is  larger than $0.535$,
and $\delta_{CP}=\pm (60^{\circ}$--$180^{\circ})$ in the region of
$\sum m_i< 160$ meV.
Since the correlation of $\sin^2\theta_{23}$ and $\delta_{CP}$ is clear,
this prediction is testable in the future experiments of neutrinos.
On the other hand,
predicted $\sin^2\theta_{12}$ and $\sin^2\theta_{13}$  cover
observed full region with $3\ \sigma$ error-bar, and there are no correlations
with $\delta_{CP}$.

The prediction of  the effective mass $\langle m_{ee}\rangle$, which is
the measure of  the neutrinoless double beta decay, is around $20$--$22$\,meV and  $45$--$60$\,meV.
We present a sample set of parameters and outputs
for  NH  in Table \ref{samplelepton}.
\begin{table}[h]
	\centering
	\begin{tabular}{|c|c|} \hline 
		\rule[14pt]{0pt}{0pt}	
		$\tau$&   $ 0.4707 + 1.2916 \, i$  \\ 
		\rule[14pt]{0pt}{0pt}
		$g_2/g_1$ &$-0.0509 + 1.2108  \, i$\\
		\rule[14pt]{0pt}{0pt}
		$\alpha_e/\gamma_e$ & $2.09\times 10^{2}$  \\
		\rule[14pt]{0pt}{0pt} 
		$\beta_e/\gamma_e$ &  $ 3.44\times 10^{3}$ \\
		\rule[14pt]{0pt}{0pt}
		$\sin^2\theta_{12}$ & $ 0.320$\\
		\rule[14pt]{0pt}{0pt}
		$\sin^2\theta_{23}$ &  $ 0.571$\\
		\rule[14pt]{0pt}{0pt}
		$\sin^2\theta_{13}$ &  $ 0.0229$\\
		\rule[14pt]{0pt}{0pt}
		$\delta_{CP}$ &  $278^\circ$ 	\\
		\rule[14pt]{0pt}{0pt}
		$\sum m_i$ &  $146$\,meV \\
		\rule[14pt]{0pt}{0pt}
		$\langle m_{ee} \rangle$ &  $22$\,meV  \\
		\rule[14pt]{0pt}{0pt}
		$\sum \chi_i^2$ &  $1.86$ 	 \\
		\hline
	\end{tabular}
	\caption{Numerical values of parameters and output
		at a sample point of NH.}
	\label{samplelepton}
\end{table} 


We have also scanned the parameter space for the case of IH of neutrino masses.
We have found parameter sets which fit the data of  $\Delta m_{\rm sol}^2$ and $\Delta m_{\rm atm}^2$
reproduce the observed  three mixing angles $\sin^2\theta_{23}$, $\sin^2\theta_{12}$, and $\sin^2\theta_{13}$.
However, the predicted $\sum m_i$ is around $200$\,meV,
which may be excluded.





It is helpful to comment on the effects of the  supersymmetry 
breaking 
and the radiative corrections because we have discussed our model
in the  limit of exact supersymmetry. 
The supersymmetry breaking effect can be neglected if
the separation between the supersymmetry breaking scale and the supersymmetry breaking mediator scale
is sufficiently large.
In our numerical results, the corrections by the renormalization are 
very small as far as we take  the relatively small value of $\tan\beta$.

\subsection{Other modular invariant flavor models}

In addition of this  $\rm \Gamma_3 \simeq A_4$ flavor model, 
other  viable models  have been also presented for  $\rm \Gamma_3 \simeq A_4$ \cite{Criado:2018thu,Ding:2019zxk,Okada:2020brs,Yao:2020qyy},
$\Gamma_4 \simeq S_4$   \cite{Novichkov:2018ovf,Kobayashi:2019mna,Wang:2019ovr,Gehrlein:2020jnr} and
for  $\Gamma_2 \simeq S_3$\cite{Kobayashi:2019rzp,Meloni:2023aru}.
The  double covering groups,  $\rm T'$~\cite{Liu:2019khw},
$\rm S_4'$ \cite{Novichkov:2020eep,Liu:2020akv}
and $\rm A_5'$ 	\cite{Wang:2020lxk,Yao:2020zml} 
have also discussed in the modular symmetry.
Subsequently these groups have been used for flavor model 
building
\cite{Okada:2022kee,Ding:2022nzn,Ding:2022aoe,Benes:2022bbg}.
Furthermore, modular forms for $\Delta(96)$ and $\Delta(384)$ have been constructed\cite{Kobayashi:2018bff},
and 
the level $7$ finite modular group $\rm \Gamma_7\simeq PSL(2,Z_7)$ as well as the level 6 group
has been  examined  for the lepton mixing\cite{Ding:2020msi,Li:2021buv}.

On the other hand, the quark mass matrix  has been also studied in the  $\rm \Gamma_3 \simeq A_4$  flavor symmetries\cite{Okada:2018yrn,Okada:2019uoy}.
Hence, the unification of  quarks and leptons  
has been applied in the framework of the  SU$(5)$ or SO$(10)$ GUT
\cite{deAnda:2018ecu,Kobayashi:2019rzp,King:2021fhl,Chen:2021zty,Du:2020ylx,Ding:2021zbg,Ding:2021eva,Ding:2022bzs}.

There are also another important physics,
the baryon asymmetry in the universe, which is
discussed with the modular symmetry.
Indeed,  the $\rm A_4$ modular flavor symmetry has been examined in
the leptogenesis\cite{Asaka:2019vev,Okada:2021qdf,Qu:2021jdy}.

The modular symmetry  keeps a residual symmetry
at the fixed points even if the modular symmetry is broken.
The $Z_3^{ST}$ symmetry generated by $ST$ remains at $\tau = \omega$, while 
the symmetry generated by $S$ remains at $\tau =i$, and it corresponds to  the $Z_4^S$ symmetry in $SL(2,Z)$ 
and the $Z_2^S$ symmetry in $PSL(2,Z)$.
Furthermore, the $Z_N^T$ symmetry in $\Gamma_N$ remains in the limit $\tau \to i \infty$.
That  gives interesting lepton mass matrices  for the flavor mixing
\cite{Novichkov:2018yse,Gui-JunDing:2019wap}.
In the modular invariant flavor model of $A_4$,  the hierarchical structure of lepton and quark flavors has been examined at nearby fixed points \cite{Okada:2020ukr}.
It is also remarked that the hierarchical structure of quark and lepton mass matrices could be derived  without fine-tuning of parameters at the nearby  fixed points of the modular symmetry
\cite{Novichkov:2021evw}. (See also Refs.\cite{Feruglio:2021dte,Feruglio:2023bav}.)
For example, the modular forms $Y_2$ and $Y_3$ among the $A_4$ triplet of weight 2 vanish 
in the limit $\tau = i \infty$.
When ${\rm Im} \tau$ is large, but finite, the $A_4$ triplet modular forms of weight 2 behave 
\begin{align}
Y_1 \sim 1, \quad Y_2 \sim \varepsilon^{1/3}, \quad Y_3 \sim \varepsilon^{2/3}
\end{align}
where $\varepsilon = e^{-2 \pi {\rm Im}\tau}$.
In general, the modular form $f(\tau)$ of $\Gamma_N$ behaves as
\begin{align}
f(\tau) \sim \varepsilon^{r/N},
\end{align}
when ${\rm Im} \tau$ is large, where $r$ denotes $Z_N^T$ charges.
These can lead to  hierarchical structures in Yukawa matrices.
Similarly, certain modular forms vanish at the fixed point $\tau = \omega$.
Around this fixed, it is convenient to define the following parameter\cite{Novichkov:2021evw}: 
\begin{align}
u \equiv \frac{\tau - \omega}{\tau - \omega^2}\ .
\end{align}
By use of this parameter, generic modular form can be approximated as 
\begin{align}
f(\tau) \sim u^r,
\end{align}
around the fixed point $\tau = \omega$, 
where $r$ depends on $Z_3^{ST}$ charges.
We have a similar behavior around the fixed pint $\tau =i$.

These behaviors around the fixed points 
allow to construct models in which 
the fermion mass hierarchies 
follow solely from the properties of the modular forms.
For example, one can derive mass matrices such as $m_{ij} \sim u^{r_i+r_j}$ and $m_{ij} \sim \varepsilon^{(r_i+r_j)/N}$ 
depending on $Z_N$ charges of matter fields.
Indeed, viable lepton and quark mass matrices 
are obtained without fine-tuning of parameters 
\cite{Novichkov:2021evw,Petcov:2022fjf,Petcov:2023vws,Kikuchi:2023cap,Abe:2023ilq,Kikuchi:2023jap,Abe:2023qmr,Abe:2023dvr}.

Further phenomenology has been developed  in many works
\cite{Kobayashi:2018wkl,Kobayashi:2019xvz,Asaka:2020tmo,Behera:2020sfe,Mishra:2020gxg,Okada:2019xqk,Kariyazono:2019ehj,Nomura:2019yft,Okada:2019lzv,Nomura:2019lnr,Criado:2019tzk,
	King:2019vhv,deMedeirosVarzielas:2020kji,Zhang:2019ngf,Nomura:2019xsb,
	Kobayashi:2019gtp,Lu:2019vgm,Wang:2019xbo,King:2020qaj,Abbas:2020qzc,
	Okada:2020oxh,Okada:2020dmb,Ding:2020yen,Nomura:2020opk,Nomura:2020cog,Okada:2020rjb,Nagao:2020azf,Nagao:2020snm,Abbas:2020vuy,
	Kuranaga:2021ujd,Okada:2021aoi,Dasgupta:2021ggp,Nomura:2021ewm,Nagao:2021rio,Nomura:2021yjb,Nomura:2021aep,Zhang:2021olk,Wang:2021mkw,
	Wang:2020dbp,Ko:2021lpx,Nomura:2021pld,Nomura:2022hxs,Otsuka:2022rak,Ding:2021iqp,
	Charalampous:2021gmf,Liu:2021gwa,Novichkov:2022wvg,Kikuchi:2022txy,Behera:2020lpd,Ahn:2022ufs,Gunji:2022xig,Abe:2023ylh,Ding:2023ynd,Kim:2023jto,Nomura:2023dgk,Kang:2022psa,deMedeirosVarzielas:2022fbw,deAnda:2023udh,Nomura:2023kwz,Ahn:2023iqa},
while theoretical investigations have been also proceeded 
\cite{Kobayashi:2016ovu,deMedeirosVarzielas:2019cyj,Baur:2020yjl,Nilles:2020nnc,Nilles:2020kgo,Nilles:2020tdp,Ding:2020zxw,Ishiguro:2020tmo,Abe:2020vmv,Kikuchi:2021yog,Baur:2021mtl,Nilles:2021ouu,Nilles:2021glx,Baur:2021bly,Baur:2022hma,	Kikuchi:2023clx,Feruglio:2023mii,Knapp-Perez:2023nty}.


\section{Texture zeros in modular symmetry}
Texture zeros of the fermion mass matrix provide an  attractive approach 
to understand the flavor mixing.  Those can be related with some flavor  symmetries.
Indeed, zeros of the  mass matrix are discussed   
in the  modular symmetry of flavors  \cite{Zhang:2019ngf,Lu:2019vgm,Ding:2022aoe}.

The flavor structure has been investigated in
magnetized orbifold models with multi-Higgs modes\cite{Abe:2008fi,Abe:2013bca,Abe:2017gye}, which are interesting compactification from higher dimensional theory such as superstring theory.
They lead to a four-dimensional chiral theory, which has the modular symmetry \cite{Kobayashi:2018rad,Kobayashi:2018bff,Ohki:2020bpo,Kikuchi:2020frp,Kikuchi:2020nxn,Kikuchi:2021ogn,Almumin:2021fbk}.
In these models, quark and lepton masses and their mixing angles were discussed  \cite{Abe:2012fj,Fujimoto:2016zjs,Abe:2014vza,Kikuchi:2021yog,Buchmuller:2017vho,Buchmuller:2017vut,Hoshiya:2022qvr}.
These magnetized orbifold models lead to multi-Higgs modes, while generic string compactification also 
leads more than one candidates for Higgs fields.

In this section, we show that texture zeros are generally  realized
at the fixed points $\tau=\omega $ in the modular flavor models with multi-Higgs \cite{Kikuchi:2022svo}.

\subsection{Quark mass matrices with mluti-Higgs}
\subsubsection{Three pairs of  Higgs fields $(1,1'')$}
We present a simple model of quark mass matrices 
in the level $N=3$ modular flavor symmetry $A_4$ 
with the multi-Higgs at $\tau=\omega$, which is referred  to as Model 1.
We assign  the $A_4$ representation and the weights $k_I$ for the relevant chiral superfields as
\begin{itemize}
	\item{quark doublet $Q=(Q^1,Q^2,Q^3)$: $A_4$ triplet with weight 2}
	\item{up-type quark singlets $(u,c,t)$: $A_4$ singlets $(1,1',1'')$ with weight 0}
	\item{down-type quark singlets $(d,s,b)$: $A_4$ singlets $(1,1',1'')$ with weight 0}
	\item{up and down sector Higgs fields $H_{u,d}^i=(H_{u,d}^1,H_{u,d}^2)$: $A_4$ singlets $(1,1'')$ with weight 0}
\end{itemize}
which are summarized in Table \ref{tab:1}.
\begin{table}[h]
	\begin{center}
		\renewcommand{\arraystretch}{1.1}
		\begin{tabular}{|c|c|c|c|c|c|} \hline
			& $Q=(Q^1,Q^2,Q^3)$ & $(u,c,t)$ & $(d,s,b)$ & $H_u$ & $H_d$ \\ \hline
			$SU(2)$ & 2 & 1 & 1 & 2 & 2 \\
			$A_4$ & 3 & $(1,1',1'')$ & $(1,1',1'')$ & $(1,1'')$ & $(1,1'')$ \\
			$k_I$ & 2 & 0 & 0 & 0 & 0 \\ \hline
		\end{tabular}
	\end{center}
	\caption{Assignments of $A_4$ representations and weights in Model 1.}
	\label{tab:1}
\end{table}

Then, the superpotential terms of the up-type quark masses and down-type quark masses are written by
\begin{align}
&W_u =
\left[\alpha^{1}_u ({\bf Y^{(2)}}Q)_1 u_{1} + \beta^{1}_u ({\bf Y^{(2)}}Q)_{1''}c_{1'} + \gamma^{1}_u ({\bf Y^{(2)}}Q)_{1'}t_{1''}\right](H_u^1)_1  \nonumber\\
&\quad + \left[\alpha^{2}_u ({\bf Y^{(2)}}Q)_{1'} u_{1} + \beta^{2}_u ({\bf Y^{(2)}}Q)_{1}c_{1'} + \gamma^{2}_u ({\bf Y^{(2)}}Q)_{1''}t_{1''}\right](H_u^2)_{1''}, \\
\nonumber\\\
&W_d =
\left[\alpha^{1}_d ({\bf Y^{(2)}}Q)_1 d_{1} + \beta^{1}_d ({\bf Y^{(2)}}Q)_{1''}s_{1'} + \gamma^{1}_d ({\bf Y^{(2)}}Q)_{1'}b_{1''}\right](H_d^1)_1 \nonumber\\\
&\quad + \left[\alpha^{2}_d ({\bf Y^{(2)}}Q)_{1'} d_{1} + \beta^{2}_d ({\bf Y^{(2)}}Q)_{1}s_{1'} + \gamma^{2}_d ({\bf Y^{(2)}}Q)_{1''}b_{1''}\right](H_d^2)_{1''},
\end{align}
where the decompositions of the tensor products are
\begin{align}
&({\bf Y^{(2)}}Q)_1 = 
\left(
\begin{pmatrix}
Y_1\\Y_2\\Y_3\\
\end{pmatrix}_3
\otimes
\begin{pmatrix}
Q^1\\Q^2\\Q^3\\
\end{pmatrix}_3
\right)_1
= Y_1Q^1+Y_2Q^3+Y_3Q^2, \\
&({\bf Y^{(2)}}Q)_{1''} = 
\left(
\begin{pmatrix}
Y_1\\Y_2\\Y_3\\
\end{pmatrix}_3
\otimes
\begin{pmatrix}
Q^1\\Q^2\\Q^3\\
\end{pmatrix}_3
\right)_{1''}
= Y_3Q^3+Y_1Q^2+Y_2Q^1, \\
&({\bf Y^{(2)}}Q)_{1'} = 
\left(
\begin{pmatrix}
Y_1\\Y_2\\Y_3\\
\end{pmatrix}_3
\otimes
\begin{pmatrix}
Q^1\\Q^2\\Q^3\\
\end{pmatrix}_3
\right)_{1'}
= Y_2Q^2+Y_1Q^3+Y_3Q^1.
\end{align}

The superpotential terms are rewritten as:
\begin{align}
W_u &= [\alpha_u^1(Y_1Q^1+Y_2Q^3+Y_3Q^2)u+\beta^1_u(Y_3Q^3+Y_1Q^2+Y_2Q^1)c\notag\\
&+\gamma^1_u(Y_2Q^2+Y_1Q^3+Y_3Q^1)t]H_u^1 \notag \\
&+[\alpha_u^2(Y_2Q^2+Y_1Q^3+Y_3Q^1)u+\beta^2_u(Y_1Q^1+Y_2Q^3+Y_3Q^2)c\notag\\
&+\gamma^2_u(Y_3Q^3+Y_1Q^2+Y_2Q^1)t]H_u^2 \notag \\
&=
\begin{pmatrix}
Q^1&Q^2&Q^3\\
\end{pmatrix}
\left(
\begin{pmatrix}
\alpha_u^1Y_1 & \beta_u^1Y_2 & \gamma_u^1Y_3 \\
\alpha_u^1Y_3 & \beta_u^1Y_1 & \gamma_u^1Y_2 \\
\alpha_u^1Y_2 & \beta_u^1Y_3 & \gamma_u^1Y_1 \\
\end{pmatrix}H_u^1
+
\begin{pmatrix}
\alpha_u^2Y_3 & \beta_u^2Y_1 & \gamma_u^2Y_2 \\
\alpha_u^2Y_2 & \beta_u^2Y_3 & \gamma_u^2Y_1 \\
\alpha_u^2Y_1 & \beta_u^2Y_2 & \gamma_u^2Y_3 \\
\end{pmatrix}H_u^2
\right)
\begin{pmatrix}
u\\c\\t\\
\end{pmatrix}, \\
W_d &= [\alpha_d^1(Y_1Q^1+Y_2Q^3+Y_3Q^2)d+\beta^1_d(Y_3Q^3+Y_1Q^2+Y_2Q^1)s\notag\\
&+\gamma^1_d(Y_2Q^2+Y_1Q^3+Y_3Q^1)b]H_d^1 \notag \\
&+[\alpha_d^2(Y_2Q^2+Y_1Q^3+Y_3Q^1)d+\beta^2_d(Y_1Q^1+Y_2Q^3+Y_3Q^2)s\notag\\
&+\gamma^2_d(Y_3Q^3+Y_1Q^2+Y_2Q^1)b]H_d^2 \notag \\
&=
\begin{pmatrix}
Q^1&Q^2&Q^3\\
\end{pmatrix}
\left(
\begin{pmatrix}
\alpha_d^1Y_1 & \beta_d^1Y_2 & \gamma_d^1Y_3 \\
\alpha_d^1Y_3 & \beta_d^1Y_1 & \gamma_d^1Y_2 \\
\alpha_d^1Y_2 & \beta_d^1Y_3 & \gamma_d^1Y_1 \\
\end{pmatrix}H_d^1
+
\begin{pmatrix}
\alpha_d^2Y_3 & \beta_d^2Y_1 & \gamma_d^2Y_2 \\
\alpha_d^2Y_2 & \beta_d^2Y_3 & \gamma_d^2Y_1 \\
\alpha_d^2Y_1 & \beta_d^2Y_2 & \gamma_d^2Y_3 \\
\end{pmatrix}H_d^2
\right)
\begin{pmatrix}
d\\s\\b\\
\end{pmatrix}.
\end{align}

Finally, the quark mass matrices  are  given as:
\begin{align}
M_u &=
\begin{pmatrix}
\alpha_u^1Y_1 & \beta_u^1Y_2 & \gamma_u^1Y_3 \\
\alpha_u^1Y_3 & \beta_u^1Y_1 & \gamma_u^1Y_2 \\
\alpha_u^1Y_2 & \beta_u^1Y_3 & \gamma_u^1Y_1 \\
\end{pmatrix} \langle H_u^1\rangle
+
\begin{pmatrix}
\alpha_u^2Y_3 & \beta_u^2Y_1 & \gamma_u^2Y_2 \\
\alpha_u^2Y_2 & \beta_u^2Y_3 & \gamma_u^2Y_1 \\
\alpha_u^2Y_1 & \beta_u^2Y_2 & \gamma_u^2Y_3 \\
\end{pmatrix} \langle H_u^2\rangle, \\
M_d &=
\begin{pmatrix}
\alpha_d^1Y_1 & \beta_d^1Y_2 & \gamma_d^1Y_3 \\
\alpha_d^1Y_3 & \beta_d^1Y_1 & \gamma_d^1Y_2 \\
\alpha_d^1Y_2 & \beta_d^1Y_3 & \gamma_d^1Y_1 \\
\end{pmatrix} \langle H_u^1\rangle
+
\begin{pmatrix}
\alpha_d^2Y_3 & \beta_d^2Y_1 & \gamma_d^2Y_2 \\
\alpha_d^2Y_2 & \beta_d^2Y_3 & \gamma_d^2Y_1 \\
\alpha_d^2Y_1 & \beta_d^2Y_2 & \gamma_d^2Y_3 \\
\end{pmatrix} \langle H_u^2\rangle,
\end{align}
where the chiralities of the mass matrix, $L$ and  $R$ are defined as $[M_{u(d)}]_{LR}$.
At the fixed point $\tau =\omega$, modular forms are given as
\begin{align}
\begin{pmatrix}
Y_1(\omega)\\Y_2(\omega)\\Y_3(\omega)
\end{pmatrix}
=Y_1(\omega)
\begin{pmatrix}
1 \\ \omega  \\ -\frac12 \omega^2
\end{pmatrix}\, .
\end{align}

\subsubsection{$ST$-eigenstate base at $\tau=\omega$}
Let us discuss the mass matrices at $\tau=\omega$ in the $ST$-eigenstates.
The $ST$-transformation of the $A_4$ triplet of the left-handed quarks $Q$ is
\begin{align}
\begin{pmatrix}
Q^1 \\ Q^2 \\ Q^3\\
\end{pmatrix}
&\xrightarrow{ST}
(-\omega-1)^{-2}
\rho(ST) 
\begin{pmatrix}
Q^1 \\ Q^2 \\ Q^3\\
\end{pmatrix}   \nonumber\\\
&= \omega^{-4}
\frac{1}{3}
\begin{pmatrix}
-1 & 2\omega & 2\omega^2 \\
2 & -\omega & 2\omega^2 \\
2 & 2\omega & -\omega^2 \\
\end{pmatrix}
\begin{pmatrix}
Q^1 \\ Q^2 \\ Q^3\\
\end{pmatrix},
\end{align}
where representations of $S$ and $T$ are given explicitly for the triplet in Appendix A.
The $ST$-eigenstate $Q'$ is obtained by using the unitary matrix $U_L$ as follows:
\begin{align}
&U_L = \frac{1}{3}
\begin{pmatrix}
2 & -\omega & 2\omega^2 \\
-\omega & 2\omega^2 & 2 \\
2\omega^2 & 2 & -\omega \\
\end{pmatrix},
\label{unitary} \\
&U_L^\dagger \omega^{-4} \rho(ST) U_L =
\begin{pmatrix}
1 & 0 & 0 \\
0 & \omega^2 & 0 \\
0 & 0 & \omega \\
\end{pmatrix}.
\end{align}
The $ST$-eigenstates are
$Q'\equiv U_L^\dagger Q$.

On the other hand,
right-handed quarks, which are singlets $(1,1',1'')$, are the eigenstates of $ST$;
that is, 
the $ST$-transformation is
\begin{align}
&\begin{pmatrix}
u \\ c \\ t \\
\end{pmatrix}
\xrightarrow{ST}
\begin{pmatrix}
1 & 0 & 0 \\
0 & \omega^2 & 0 \\
0 & 0 & \omega \\
\end{pmatrix}
\begin{pmatrix}
u \\ c \\ t \\
\end{pmatrix}, \quad
\begin{pmatrix}
d \\ s \\ b \\
\end{pmatrix}
\xrightarrow{ST}
\begin{pmatrix}
1 & 0 & 0 \\
0 & \omega^2 & 0 \\
0 & 0 & \omega \\
\end{pmatrix}
\begin{pmatrix}
d \\ s \\ b \\
\end{pmatrix}.
\end{align}

Higgs fields are also the $ST$-eigenstates since they are singlets $(1,1'')$.
Therefore, $ST$-transformation of them is
\begin{align}
&\begin{pmatrix}
H_{u,d}^1 \\ H_{u,d}^2 \\
\end{pmatrix}
\xrightarrow{ST}
\begin{pmatrix}
1 & 0 \\
0 & \omega \\
\end{pmatrix}
\begin{pmatrix}
H_{u,d}^1 \\ H_{u,d}^2 \\
\end{pmatrix}.
\end{align}

In the $ST$-eigenstates,  the quark mass matrices are
transformed by Eq.\,\eqref{unitary}. It is  given as:
\begin{align}
&U_L^T M_u = c
\begin{pmatrix}
\alpha_u^1 & 0 & 0 \\
0 & 0 & \gamma_u^1 \\
0 & \beta_u^1 & 0 \\
\end{pmatrix}\langle H_u^1\rangle
+ c
\begin{pmatrix}
0 & \beta_u^2 & 0 \\
\alpha_u^2 & 0 & 0 \\
0 & 0 & \gamma_u^2 \\
\end{pmatrix}\langle H_u^2\rangle, 
\label{quark-mass-matrix-up}\\
&U_L^T M_d = c
\begin{pmatrix}
\alpha_d^1 & 0 & 0 \\
0 & 0 & \gamma_d^1 \\
0 & \beta_d^1 & 0 \\
\end{pmatrix}\langle H_d^1\rangle
+ c
\begin{pmatrix}
0 & \beta_d^2 & 0 \\
\alpha_d^2 & 0 & 0 \\
0 & 0 & \gamma_d^2 \\
\end{pmatrix}\langle H_d^2\rangle,
\label{quark-mass-matrix-down}
\end{align}
where $c=\sqrt{|Y_1|^2+|Y_2|^2+|Y_3|^2}$.
Thus, some zeros appear for quark mass matrices.

Now,  we impose $\alpha^1_{u,d}=0$, we obtain the  nearest neighbor interaction (NNI) form,\footnote{The NNI form of three families  has vanishing entries of (1,1),\, (2,2),\, (1,3),\, (3,1), but
	is  not  necessary to be Hermitian.}
which is considered as a general form of both
up- and down-types quark mass matrices because this form is achieved by the transformation that leaves the left- handed gauge interaction invariant
\cite{Branco:1988iq}. 
The NNI form is a desirable base  to derive the Fritzsch-type quark mass matrix
while the NNI form   is  a general form of quark mass matrices.

Therefore, the quark masses and the CKM matrix are reproduced taking relevant values
of parameters.
It is noticed that the flavor mixing is not realized in the case of one Higgs doublets for up- and down-type quark sectors. 
Thus, the NNI forms at $\tau=\omega$  are simply obtained 
unless the VEVs of two-Higgs vanish.

The CP symmetry is not violated at $\tau = \omega$ in modular flavor symmetric models with 
a pair of Higgs fields because of the $T$ symmetry \cite{Kobayashi:2019uyt}.
However, the models with multi-Higgs fields can break the CP symmetry at 
the fixed point $\tau = \omega$ even if all of the Higgs VEVs are real \cite{Kikuchi:2022geu}.
Thus,  the CP phase appears in our  models, in general.
Our models are interesting from the viewpoint of the CP violation, too.


The non-vanishing VEVs of both Higgs fields $H^1_{u,d}$ and $H^2_{u,d}$ are important to 
realize the NNI forms.
We expect the scenario that these Higgs fields have a $\mu$-matrix to 
mix them,
\begin{align}
W_\mu = \mu_{ij}H^i_u H^j_d .
\end{align}
Then, a light linear combination develops its VEV, which includes 
$H^1_{u,d}$ and $H^2_{u,d}$.
However, the above assignment of $A_4$ representations $(1,  1'')$ for the Higgs fields allows 
the $\mu$-term of only $\mu_{11}$, and the others vanish.
That is, the mixing does not occur.
When we assume the singlet $S$ with the $A_4$ $1'$ representation develops its VEV, 
the $(1,2)$ and $(2,1)$ elements appear as $\mu_{12}=\mu_{21}=\lambda \langle S \rangle$ 
like the next-to-minimal supersymmetric standard model.

It is noted that the  alternative assignment of weights for  the Higgs and the left-handed quarks 
also gives desirable $\mu$ term \cite{Kikuchi:2022svo}.


\subsubsection{Three pairs of  Higgs fields $(1,1'',1')$}

We also study three pairs of Higgs fields with the $A_4$ $(1,1'',1')$ representations.
We add another pair  of Higgs fields $H^3_{u,d}$ with the $A_4$ $1'$ representation of the modular weight $0$.
Then, we easily  obtain  the mass matrices  as follows:
\begin{align}
&U_L^T M_u = c
\begin{pmatrix}
\alpha_u^1 & 0 & 0 \\
0 & 0 & \gamma_u^1 \\
0 & \beta_u^1 & 0 \\
\end{pmatrix}\langle H_u^1\rangle
+ c
\begin{pmatrix}
0 & \beta_u^2 & 0 \\
\alpha_u^2 & 0 & 0 \\
0 & 0 & \gamma_u^2 \\
\end{pmatrix}\langle H_u^2\rangle
+ c
\begin{pmatrix}
0 & 0 & \gamma_u^3 \\
0 & \beta_u^3 & 0 \\
\alpha_u^3 & 0 & 0 \\
\end{pmatrix}\langle H_u^3\rangle, \\
&U_L^T M_d = c
\begin{pmatrix}
\alpha_d^1 & 0 & 0 \\
0 & 0 & \gamma_d^1 \\
0 & \beta_d^1 & 0 \\
\end{pmatrix}\langle H_d^1\rangle
+ c
\begin{pmatrix}
0 & \beta_d^2 & 0 \\
\alpha_d^2 & 0 & 0 \\
0 & 0 & \gamma_d^2 \\
\end{pmatrix}\langle H_d^2\rangle
+ c
\begin{pmatrix}
0 & 0 & \gamma_d^3 \\
0 & \beta_d^3 & 0 \\
\alpha_d^3 & 0 & 0 \\
\end{pmatrix}\langle H_d^3\rangle.
\end{align}
This model can lead to a quite generic mass matrix.
For example, by setting some of $\alpha^i_{u,d}, \beta^i_{u,d}, \gamma^i_{u,d}$ to be zero, 
we can drive some of texture zero structures including the NNI form.
In addition, we can assume $\beta^i_{u,d}=\gamma^i_{u,d}$ or $\beta^i_{u,d}=(\gamma^i_{u,d})^*$ 
to reduce the number of free parameters and realize a certain form of mass matrices.
Thus, the different assignment of the $A_4$ singlets $(1,1'',1')$  for Higgs leads to different 
texture zeros.


\subsection{Extensions  of  models}

In this section,  the quark mass matrices are discussed in the specific modular symmetry
of $N=3$ in order to show the derivation of NNI forms clearly.

It is noted that one can obtain flavor  models leading to the NNI forms 
in the  $S_4$ and $A_5$ modular flavor symmetries.
Such  texture zero structure originates from  the  $ST$ charge of the residual symmetry $Z_3$ of $SL(2,Z)$.
The NNI form can be realized at the fixed point $\tau = \omega$ in $A_4$ and $S_4$ modular flavor models with two pairs of Higgs fields,
when we assign properly modular weights to Yukawa couplings and $A_4$ and $S_4$ representations to 
three generations of quarks.
It is found that four pairs of Higgs fields to realize the NNI form in $A_5$ modular flavor models.
Thus, the modular flavor models with multi-Higgs fields at the fixed point $\tau = \omega$
leads to successful  quark mass matrices \cite{Kikuchi:2022svo}.

Texture zeros have been studied phenomenologically in the lepton sector 
\cite{Fukugita:2003tn,Xing:2004xu,Obara:2006ny,Fukugita:2012jr,
	Fukugita:2016hzf,Fritzsch:2012rg}.
We can extend our formula  of the quark mass matrices to the lepton sector.
Extension to the charged lepton mass matrix is straightforward, 
and we obtain the same results.
On the other hand, there is some freedoms for the neutrino mass matrix,
depending on the mechanism of producing  tiny masses, for example,  seesaw mechanism. 

\section{CP Symmetry}

In this section, we study CP violation in modular symmetric flavor models.


The 4D CP symmetry can be embedded into a proper Lorentz transformation in a higher dimensional 
theory.
Here, we concentrate on 6D theory, that is, extra two dimensions in addition to our 4D space-time.
$T^2$ is one of examples of two-dimensional compact space.
We denote the coordinate on extra dimension, e.g. $T^2$, by $z$.
Then, we consider the following transformation 
\begin{align}
z \to -z^*,
\end{align}
at the same time as the 4D CP transformation.
Such a combination is included in a 6D proper Lorentz symmetry.
Because of the above coordinate transformation, the modulus $\tau$ on $T^2$ 
transforms
\begin{align}
\tau \to -\tau^*,
\label{eq:CP-tau}
\end{align}
under the CP symmetry \cite{Baur:2019kwi,Novichkov:2019sqv}.
Note that the upper half plane of $\tau$ maps onto itself by this transformation.
Another transformation such as $z \to z^*$ can also correspond to a 6D proper Lorentz symmetry, 
but such a transformation maps the  upper half plane onto the lower half plane.
Thus, we do not use such a transformation.

Obviously, we find that the line ${\rm Re}\tau = 0$ is CP invariant.
Other values are also CP invariant up to the modular symmetry.
For example, $\tau=\omega=e^{2\pi i/3}$ transforms
\begin{align}
\tau = \omega = \frac{-1 + \sqrt{3} i }{2} \to -\tau^*=\frac{1 + \sqrt{3} i }{2} ,
\end{align}
under the CP transformation Eq.\,(\ref{eq:CP-tau}).
However, these values are related with each other by the $T$-transformation.
Thus, the fixed point $\tau=\omega$ is also CP invariant point.
Similarly, the line ${\rm Re}\tau = \pm 1/2$ is CP invariant.


The typical K\"ahler potential of the modulus field $\tau$ is written by 
\begin{align}
K=-\ln [2{\rm Im}\tau].
\end{align}
The K\"ahler potential is invariant under the transformation, $\tau \to -\tau^*$.
In addition, the superpotential $|\hat W|^2$ is invariant if it transforms
\begin{align}
W(\tau) \to W(-\tau^*)=e^{i\chi}\overline{W(\tau)},
\end{align}
under the CP symmetry with $\tau \to -\tau^*$ including the CP transformation of chiral matter fields.

We study the CP violation through the modulus stabilization.
One of the moduli stabilization scenarios is due to the three-form fluxes \cite{Gukov:1999ya}.
Indeed, the moduli stabilization due to the three-from fluxes was studied in modular flavor models 
in Ref.~\cite{Ishiguro:2020tmo}.
Its result shows that the fixed point $\tau = \omega$ is favored statistically with highest probability.
The above discussions implies that the CP violation does not occur at this fixed point.
In Ref.~\cite{Abe:2020vmv} the moduli stabilization was studied by one-loop induced Fayet-Illiopoulos terms, and 
the modulus $\tau$ is stabilized at the same fixed point\footnote{See also for recent studies on moduli stabilization in 
	modular flavor models Refs.~\cite{Novichkov:2022wvg,Ishiguro:2022pde}.}.
In addition, we study another mechanism of the moduli stabilization by assuming non-perturbative effects.
We start with the superpotential $W=m(\tau)Q \bar Q$ with the $A_4$ modular flavor symmetry.
Then, we assume the condensation $\langle Q \bar Q\rangle \neq 0$.
The superpotential is $A_4$ trivial singlet.
We assume the following superpotential:
\begin{align}
W= \Lambda_dY^{(4)}_{\bf 1}(\tau),
\end{align}
where $\Lambda_d$ corresponds to $\langle Q \bar Q\rangle$ and must have a proper modular weight.
The minimum of the supergravity scalar potential with the above superpotential is obtained as 
$\tau_{min}=1.09\,i + p/2$,
where $p$ is odd integer\cite{Kobayashi:2019uyt}.
The above discussion implies that the CP violation does not occur at this point.
On the other hand, we assume the following superpotential:
\begin{align}
W= \Lambda_d(Y^{(4)}_{\bf 1}(\tau))^{-1},
\end{align}
where $\Lambda_d$ must have a proper modular weight.
The minimum of the supergravity scalar potential with the above superpotential is obtained as 
$\tau_{min}=1.09\,i + n$,
where $n$ is integer\cite{Kobayashi:2019uyt}.
Obviously, this is CP invariant point.
Similarly, we can study other modular flavor models such as $S_3$ and $S_4$ modular symmetries, 
and the potential minimum corresponds to either ${\rm Re}\,\tau = 0$ or $1/2$ (mod 1) \cite{Kobayashi:2019uyt}.
In both cases, CP violation does not occur.

We examine explicitly mass matrices at ${\rm Re}\,\tau = 0$ and $1/2$ 
in order to understand that the CP symmetry is not violated at these lines.
We study the flavor model with the $\Gamma_N$ modular flavor symmetry.
We use the basis that $\rho(T)$ is diagonal and satisfies $\rho(T)^N=1$.
Then, the chiral fields $\Phi_i$ such as left-handed quarks $Q_i$, up-sector and down-sector right-handed quarks $u_i$, 
$d_i$, and the Higgs field $H_{u,d}$ as well as lepton fields 
transform 
\begin{align}
\Phi_i \to e^{2\pi i P[\Phi_i]/N }\Phi_i,
\end{align}
under the $T$-transformation, where $P[\Phi_i]$ is integer.
That is the $Z_N^{(T)}$ rotation.
Here, we assume one pair of Higgs fields $H^u$ and $H^d$, which are trivial singlets under the $\Gamma_N$ modular symmetry.
Then, the quark Yukawa terms in the superpotential can be written by
\begin{align}
\hat W = Y^{(u)}_{ij}(\tau) Q_iu_jH^u + Y^{(d)}_{ij}Q_i d_j H^d.
\end{align}
We replace the Higgs fields by their VEVs so as to obtain the mass terms,
\begin{align}
\hat W=M^u_{ij}(\tau) Q_i u_j  + M^d_{ij}(\tau) Q_i d_j .
\end{align}
Note that the Yukawa couplings are modular forms.
Then, the above mass matrices can also be written by modular forms after replacing the Higgs fields by their VEVs.
Since these mass terms must be invariant under the $T$-transformation,
the mass matrix must transform as
\begin{align}
M^u_{ij}(\tau) \to e^{-2\pi i(P[Q_i]+P[u_j])/N}M^u_{ij}, 
\qquad M^d_{ij}(\tau) \to e^{-2\pi i(P[Q_i]+P[d_j])/N}M^d_{ij}.
\end{align}
That implies that the mass matrices can be written by
\begin{align}
M^u_{ij}(\tau)=m^u_{ij}(q)q^{-(P[Q_i]+P[u_j])/N}= m^u_{ij}(q)e^{-2\pi i(P[Q_i]+P[u_j])\tau/N}, \notag \\
M^d_{ij}(\tau)=m^d_{ij}(q)q^{-(P[Q_i]+P[d_j])/N}= m^d_{ij}(q)e^{-2\pi i(P[Q_i]+P[d_j])\tau/N}, 
\end{align}
in terms of $q=e^{2\pi i \tau}$, where $m^{u,d}_{ij}(q)$ include series of integer powers of $q$ as
\begin{align}
m^{u,d}_{ij}(q)=a_0^{u,d}+a_1^{u,d}q+a_2^{u,d}q^2+\cdots.
\end{align}

It is obvious that all of the entries in $M^{u,d}_{ij}(\tau)$ are real when ${\rm Re}\,\tau=0$.
CP is not violated.
On the other hand, it seems that the mass matrix has phases for other values of ${\rm Re}\,\tau$.
For example, when ${\rm Re }\,\tau = 1/2$, 
the phase structure of the mass matrix can be written by
\begin{align}
M^u_{ij}=\tilde m^u_{ij}e^{-\pi i(P[Q_i]+P[u_j])/N}, \qquad M^d_{ij}=\tilde m^d_{ij}e^{-\pi i(P[Q_i]+P[d_j])/N}, 
\end{align}
where $\tilde m^u_{ij}=m^u_{ij}e^{-2\pi (P[Q_i]+P[u_j]){\rm Im}\tau /N}$,  
$\tilde m^d_{ij}=m^d_{ij}e^{-2\pi (P[Q_i]+P[d_j]){\rm Im}\tau /N}$, and they are real.
However, such phases can be canceled by rephasing 
\begin{align}
\Phi_i \to \Phi e^{\pi i P[\Phi_i]/N}\Phi_i,
\end{align}
and there is no physical CP phase for 
${\rm Re}\, \tau =1/2$.
That is the $Z_{2N}^{(T)}$ rotation.
Note that $m_{ij}(q)$ can have a physical CP phase, which can not be  canceled, 
except ${\rm Re}\,\tau =0, 1/2$.
Similarly, we can discuss the lepton sector, and 
the CP phase does not appear when ${\rm Re}\,\tau =0, 1/2$.

The fixed point $\tau = \omega$ is statistically favored with highest probability, and 
phenomenological interesting because there remains $Z_3$ symmetry.
However, the CP violation does not occur in modular flavor models with one pair of Higgs fields.
That suggests extension to models with multi-Higgs fields.
Indeed many string compactifications lead more than one candidates of Higgs fields, which have 
the same quantum numbers of $SU(3)\times SU(2) \times U(1)_Y$ and can couple with quarks and leptons.
We extend the above discussion to modular flavor models with multi-Higgs fields $H^{u,d}_\ell$.
The quark Yukawa terms in the superpotential can be written by
\begin{align}
\hat W = Y^{(u)}_{ij\ell}(\tau) Q_iu_jH^u_\ell + Y^{(d)}_{ij\ell}Q_i d_j H^d_\ell.
\end{align}
Since these terms are invariant under the $T$-transformation, 
Yukawa couplings must transform as
\begin{align}
Y^{(u)}_{ij\ell}(\tau)  \to e^{2 \pi i P[Y^u_{ij\ell}]}Y^{(u)}_{ij\ell}(\tau), 
\qquad Y^{(d)}_{ij\ell}(\tau)  \to e^{2 \pi i P[Y^d_{ij\ell}]}Y^{(d)}_{ij\ell}(\tau), 
\end{align}
under the $T$-transformation, 
where
\begin{align}
P[Y^u_{ij\ell}]=-(P[Q_i]+P[u_j]+P[H^u_\ell]), \qquad P[Y^d_{ij\ell}]=-(P[Q_i]+P[d_j]+P[H^d_\ell]).
\label{eq:Yukawa-Tcharge}
\end{align}
That implies that the modular forms of Yukawa couplings can be written by 
\begin{align}
Y^{(u)}_{ij\ell}(\tau)& =a_0q^{P[Y^u_{(i j \ell)}]/N} + a_1 qq^{P[Y^u_{(i j \ell)}]/N} +a_2 q^2q^{P[Y^u_{(i j \ell)}]/N}
+ \cdots \nonumber\\
&=\tilde Y^{(u)}_{ij\ell}(q) q^{P[Y^u_{(i j \ell)}]/N}, \notag \\
Y^{(d)}_{ij\ell}(\tau) &=b_0q^{P[Y^d_{(i j \ell)}]/N} + b_1 qq^{P[Y^d_{(i j \ell)}]/N} +b_2 q^2q^{P[Y^d_{(i j \ell)}]/N} 
+ \cdots \nonumber\\
&=\tilde Y^{(d)}_{ij\ell}(q) q^{P[Y^u_{(i j \ell)}]/N},
\end{align}
where the functions $\tilde Y^{(u)}_{ij\ell}(q) $ and $\tilde Y^{(d)}_{ij\ell}(q) $ are series of positive integer powers of $q$.

We denote Higgs VEVs by
\begin{align}
v^u_\ell =|v^u_\ell|e^{2\pi iP[v^u_\ell]/N}=\langle H^u_\ell \rangle, \qquad 
v^d_\ell =|v^d_\ell|e^{2\pi iP[v^d_\ell]/N}=\langle H^d_\ell \rangle,
\end{align}
where $P[v^u_\ell]$ or $P[v^d_\ell]$ is not integer for a generic VEV.
Then, the mass matrices can be written by 
\begin{align}
M^{u,d}_{ij}=\sum_\ell Y^{(u,d)}_{ij \ell} v^{u,d}_\ell.
\end{align}
When ${\rm Re}\,\tau=0$, all of the Yukawa coupligs $Y^{(u,d)}_{ij \ell}$ are real.
In this case, the non-trivial CP phase appears only if the VEVs $v^{u,d}_\ell$ have 
phases different relatively from each other.
When ${\rm Re}\,\tau=-1/2$, e.g. $\tau = \omega$, the Yukawa coupligs $Y^{(u,d)}_{ij \ell}$ have different phases.
Thus, the non-trivial CP phase appears for generic values of VEVs.
However, if they satisfy 
\begin{align}
&-\frac12 \left(P[Y^{u}_{(ij\ell)}] + P[{Q_i}]+P[{u_j}] \right) + P[v^{u}_\ell]= {\rm constant~independent~of}~\ell, \notag \\
&-\frac12 \left(P[Y^{d}_{(ij\ell)}] + P[{Q_i}]+P[{d_j}] \right) + P[v^{d}_\ell]= {\rm constant~independent~of}~\ell,
\label{eq:CP-condition}
\end{align} 
for all of allowed Yukawa couplings with $i,j$ fixed, 
one can cancel phases of mass matrix elements up to an overall phase by $Z_{2N}^{(T)}$ rotation.
We can compare this condition with the relations Eq.\,(\ref{eq:Yukawa-Tcharge}), where the factor $-1/2$ originates from 
${\rm Re}\,\tau=-1/2$.
Thus, the $T$-symmetry determines the VEV direction $v^{u,d}_\ell$, where the CP symmetry remains.
CP violation was also studied in an explicit magnetized orbifold model \cite{Kikuchi:2022geu}.

\section{SMEFT}

So far, we have studied renormalizable coupligs such as Yukawa couplings.
Since the SM is effective theory of underlying theory, it may include 
higher dimensional operators and they may lead to flavor and CP violating processes. 
Here, we study higher dimensional operators.

The SM with renormalizable couplings has the $U(3)^5$ flavor symmetry 
in the limit that all of the Yukawa couplings vanish, 
where the $U(3)^5$ symmetry is explicitly written by 
$U(3)_Q\times U(3)_u \times U(3)_d \times U(3)_L \times U(3)_e$ and they correspond to the 
symmetries of three generations of left-handed quarks, up-sector and down-sector right-handed quarks, left-handed leptons, and  
right-handed charged leptons.
Even for non-vanishing Yukawa couplings, 
the SM can have the   $U(3)^5$ flavor symmetry by assuming that 
Yukawa couplings are spurion fields, which  transform non-trivially under the $U(3)^5$ flavor symmetry.
That is, the up-sector and down-sector Yukawa couplings transform as 
$({\bf 3}, \bar{\bf 3},{\bf 1},{\bf 1},{\bf 1})$ and $({\bf 3}, {\bf 1},\bar{\bf 3},{\bf 1},{\bf 1})$
under the symmetry $U(3)_Q\times U(3)_u \times U(3)_d \times U(3)_L \times U(3)_e$ 
while the lepton Yukawa couplings transform as  
$({\bf 1},{\bf 1},{\bf 1},{\bf 3}, \bar{\bf 3})$.
We require that higher dimensional operators also satisfy the  $U(3)^5$ flavor symmetry.
Then coefficients of higher dimensional operators can be written in terms of Yukawa couplings, which are 
spurion fields.
That is the MFV scenario \cite{Chivukula:1987py,DAmbrosio:2002vsn}.

We can compute $n$-point couplings within the framework of superstring theory.
For example, 
$n$-point couplings were calculated in intersecting D-brane models \cite{Cvetic:2003ch,Abel:2003vv,Abel:2003yx}, 
magnetized D-brane models\cite{Cremades:2004wa,Abe:2009dr}, and heterotic orbifold models 
\cite{Hamidi:1986vh,Dixon:1986qv,Atick:1987kd,Burwick:1990tu,Stieberger:1992bj,Choi:2007nb}.
These computations are carried out by two-dimensional conformal field theory (CFT) and 
integral of products of wave functions in compact space.

Massless modes in string theory correspond to vertex operators $V_i(z)$ in CFT, where 
$w$ denotes the complex coordinate on the world-sheet.
These vertex operators satisfy the operator product expansion,
\begin{align}
V_i (w) V_j (0) \sim \sum_k \frac{y_{ijk}}{w^{h_k -h_i -h_j}}V_k(0),
\end{align}
where $h_i$ denote the conformal dimensions of vertex operators $V_i$.
The coefficients $y_{ijk}$ provide us with 3-point couplings among massless modes corresponding 
vertex operators, $V_i$, $V_j$, $V_k$ in low-energy effective field theory.
Furthermore, 4-point couplings $y_{ijk\ell}$ can be written by products of 3-point couplings,
\begin{align}
y_{ijk\ell}=\sum_m y_{ijm}y_{mk\ell}.
\label{eq:stringy-Ansatz}
\end{align}
Similarly, generic $n$-point couplings can be written by products of 3-point couplings.
That implies that when the 3-point couplings $y_{ijk}$ have the modular symmetry, 
4-point couplings and higher order couplings are also controlled by the modular symmetry.
Indeed, these couplings can be written by modular forms, which depend on 
the moduli fields.
In this sense, these couplings are spurion fields.
Thus, this theory can provide us with the stringy origin of minimal flavor violation, 
where the flavor symmetry is the modular symmetry instead of $U(3)^5$.


Similarly, various classes of 4D low-energy effective field theory derived 
string theory satisfy the requirement of minimal flavor violation hypothesis 
at the compactification scale.
However, several physical stages may occur between the compactification scale 
and low energy scale, 
(i) some modes gain masses and (ii) some scalar fields develop their VEVs.
At the stage (i), we just integrate out massive modes.
Effective field theory after such an integration also satisfies the above structure.
At the stage (ii), new operators appear.
For example, suppose that we have the coupling, 
$y_{ijk\ell} \phi_i \phi_j \phi_k \phi_\ell$ and $\phi_i$ develops its VEV.
Then, the new operator $y'_{jk\ell} \phi_j \phi_k \phi_\ell$ appears, where 
$y'_{jk\ell}=y_{ijk\ell}\langle \phi_i \rangle$.
Both $y'_{jk\ell}$ and $y_{ijk\ell}$ are spurion fields, and 
the transformation behavior of $y'_{jk\ell}$ is the same as $y_{ijk\ell}\langle \phi_i \rangle$.
Thus, the minimal flavor violation structure with the modular symmetry 
is not violated.

One of non-trivial symmetry breaking is the supersymmetry breaking.
The supersymmetry breaking can occur by non-vanishing F-terms.
If all of the F-terms are trivial singlets under the modular symmetry, 
obviously all of the soft terms are modular invariant.
The supersymmetry breaking due to modulus F-term is non-trivial 
from the viewpoint of the modular symmetry.
Detailed study was done in Ref.~\cite{Kikuchi:2022pkd}.
It was found that all of the soft terms except the B-term are 
modular invariant.
If the generation mechanism of the $\mu$-term is modular invariant, 
the B-term is also modular invariant.

If the above scenario holds true, the low-energy effective field theory 
around the weak scale has the minimal flavor violation structure with the 
modular symmetry.
That is, the SMEFT can have the modular symmetry.
For example, there appear the four-fermi operators and dipole operators,
\begin{align}
\frac{y_{ijk\ell}}{\Lambda^2} (\bar \Psi_i \Gamma \Psi_j)(\bar \Psi_k \Gamma \Psi_\ell), 
\qquad 
\frac{c_{ij}v}{\Lambda^2} (\bar \Psi_i \sigma^{\mu \nu} \Psi_j)F_{\mu \nu},
\end{align}
where $\Gamma$ denotes a generic combination of gamma matrices.
These operators must be modular invariant and their coefficients $y_{ijk\ell}$ and $c_{ij}$ are modular forms.
Furthermore, the coefficients $y_{ijk\ell}$ can be written by 
productions of 3-point couplings as Eq.\,(\ref{eq:stringy-Ansatz}), 
where the mode $m$ may correspond to known modes like the Higgs field or unknown modes.
The cut-off scale $\Lambda$ depends on the scenario with the stages (i) and (ii), that is, 
mass scales and symmetry breaking scales including the supersymmetry breaking scale.
Phenomenological implications of modular symmetric SMEFT were 
studied, e.g. flavor violations and lepton $(g-2)$ processes \cite{Kobayashi:2021uam,Kobayashi:2021pav,Kobayashi:2022jvy}.

\section{Conclusion}

We have reviewed on modular flavor symmetric models from several viewpoints, 
realization of fermion mass matrices, the texture structure, the CP violation and higher dimensional operators in SMEFT.
Indeed many works have been done recently, in particular in realization of quark and lepton masses and mixing angles 
as well as the CP violation. 
In addition, the modular flavor symmetry have been used  for dark matter, inflation models, and leptogenesis in bottom-up approach.
The modular flavor symmetry may originate from compactification of higher dimensional theory such as superstring theory.
Also the modular flavor symmetry have been studied in top-down approach.
Thus, the modular flavor symmetry can become a bridge to connect the low-energy physics and high-energy physic such as 
superstring theory and would provide us with a missing piece to solve the flavor puzzle in particle physics.

\section*{Acknowledgement}

The authors would like to thank Y.~Abe, T.~Higaki, K.~Ishiguro, J.~Kawamura, S.~Kikuchi, S.~Nagamoto,   K.~Nasu, T.~Nomura, H.~Okada, N.~Omoto, Y.~Orisaka, H.~Otsuka, S.T.~Petcov, Y.~Shimizu, T.~Shimomura, S.~Takada, K.~Takagi, S.~Tamba, K.~Tanaka, T.H.~Tatsuishi, H.~Uchida, S.~Uemura, K.~Yamamoto, T.~Yoshida for useful discussions.

\appendix
\section*{Appendix}

\section{Modular forms of $A_4$}
	\label{Modularforms}
	%
	%
	The  modular forms of weight $2$ transforming
	as a triplet of $A_4$ can be written in terms of 
	$\eta(\tau)$ and its derivative \cite{Feruglio:2017spp}:
	\begin{eqnarray} 
	\label{weight2}
	Y_1 &=& \frac{i}{2\pi}\left( \frac{\eta'(\tau/3)}{\eta(\tau/3)}  +\frac{\eta'((\tau +1)/3)}{\eta((\tau+1)/3)}  
	+\frac{\eta'((\tau +2)/3)}{\eta((\tau+2)/3)} - \frac{27\eta'(3\tau)}{\eta(3\tau)}  \right), \nonumber \\
	Y_2 &=& \frac{-i}{\pi}\left( \frac{\eta'(\tau/3)}{\eta(\tau/3)}  +\omega^2\frac{\eta'((\tau +1)/3)}{\eta((\tau+1)/3)}  
	+\omega \frac{\eta'((\tau +2)/3)}{\eta((\tau+2)/3)}  \right) , \label{Yi} \\ 
	Y_3 &=& \frac{-i}{\pi}\left( \frac{\eta'(\tau/3)}{\eta(\tau/3)}  +\omega\frac{\eta'((\tau +1)/3)}{\eta((\tau+1)/3)}  
	+\omega^2 \frac{\eta'((\tau +2)/3)}{\eta((\tau+2)/3)}  \right)\,,
	\nonumber
	\end{eqnarray}
	%
	which satisfy also the constraint \cite{Feruglio:2017spp}:
	\begin{align}
	Y_2^2+2Y_1Y_3=0~.
	\label{condition}
	\end{align}
	%
	They have the following  $q$-expansions:
	\begin{align}
	{\bf Y^{(2)}_3}
	=\begin{pmatrix}Y_1\\Y_2\\Y_3\end{pmatrix}=
	\begin{pmatrix}
	1+12q+36q^2+12q^3+\dots \\
	-6q^{1/3}(1+7q+8q^2+\dots) \\
	-18q^{2/3}(1+2q+5q^2+\dots)\end{pmatrix}\,,
	\label{Y(2)}
	\end{align}
	%
	where
	\begin{align}
	q=\exp{ (2\pi i\,\tau)}\,.
	\label{q}
	\end{align}
	%
	
	The five modular forms of weight 4 are given as:
	\begin{align}
	&\begin{aligned}
	{\bf Y^{(\rm 4)}_1}=Y_1^2+2 Y_2 Y_3 \ , \quad
	{\bf Y^{(\rm 4)}_{1'}}=Y_3^2+2 Y_1 Y_2 \ , \quad
	{\bf Y^{(\rm 4)}_{1''}}=Y_2^2+2 Y_1 Y_3=0 \ , \quad
	\end{aligned}\nonumber \\
	\nonumber \\
	&\begin{aligned} {\bf Y^{(\rm 4)}_{3}}=
	\begin{pmatrix}
	Y_1^{(4)}  \\
	Y_2^{(4)} \\
	Y_3^{(4)}
	\end{pmatrix}
	=
	\begin{pmatrix}
	Y_1^2-Y_2 Y_3  \\
	Y_3^2 -Y_1 Y_2 \\
	Y_2^2-Y_1 Y_3
	\end{pmatrix}\ , 
	\end{aligned}
	\label{weight4}
	\end{align}
	where ${\bf Y^{(\rm 4)}_{1''}}$ vanishes due to the constraint of Eq.\,(\ref{condition}).
	%
	
	For weigh 6, there are  seven modular forms as:
	\begin{align}
	&\begin{aligned}
	{\bf Y^{(\rm 6)}_1}=Y_1^3+ Y_2^3+Y_3^3 -3Y_1 Y_2 Y_3  \ , 
	\end{aligned} \nonumber \\
	\nonumber \\
	&\begin{aligned} {\bf Y^{(\rm 6)}_3}\equiv 
	\begin{pmatrix}
	Y_1^{(6)}  \\
	Y_2^{(6)} \\
	Y_3^{(6)}
	\end{pmatrix}
	=(Y_1^2+2 Y_2 Y_3)
	\begin{pmatrix}
	Y_1  \\
	Y_2 \\
	Y_3
	\end{pmatrix} , \quad
	\end{aligned}
	\begin{aligned} {\bf Y^{(\rm 6)}_{3'}}\equiv
	\begin{pmatrix}
	Y_1^{'(6)}  \\
	Y_2^{'(6)} \\
	Y_3^{'(6)}
	\end{pmatrix}
	=(Y_3^2+2 Y_1 Y_2 )
	\begin{pmatrix}
	Y_3  \\
	Y_1 \\
	Y_2
	\end{pmatrix} . 
	\end{aligned}
	\label{weight6}
	\end{align}
	For weigh 8, there are  nine modular forms as:
	\begin{align}
	&\begin{aligned}
	{\bf Y^{(\rm 8)}_1}=(Y_1^2+2Y_2 Y_3)^2  , \ \
	{\bf Y^{(\rm 8)}_{1'}}=(Y_1^2+2Y_2 Y_3)(Y_3^2+2Y_1 Y_2) , \ \ 
	{\bf Y^{(\rm 8)}_{1"}}=(Y_3^2+2Y_1 Y_2)^2 , \quad 
	\end{aligned} \nonumber \\
	\nonumber \\
	&\begin{aligned} {\bf Y^{(\rm 8)}_3}\equiv 
	\begin{pmatrix}
	Y_1^{(8)}  \\
	Y_2^{(8)} \\
	Y_3^{(8)}
	\end{pmatrix} 
	=(Y_1^2+2 Y_2 Y_3) 
	\begin{pmatrix}
	Y_1^2-Y_2 Y_3  \\
	Y_3^2 -Y_1 Y_2 \\
	Y_2^2-Y_1 Y_3
	\end{pmatrix} , \ 
	\end{aligned}\nonumber\\
	&\begin{aligned} {\bf Y^{(\rm 8)}_{3'}}\equiv 
	\begin{pmatrix}
	Y_1^{'(8)}  \\
	Y_2^{'(8)} \\
	Y_3^{'(8)}
	\end{pmatrix} 
	=(Y_3^2+2 Y_1 Y_2 ) 
	\begin{pmatrix}
	Y_2^2-Y_1 Y_3 \\
	Y_1^2-Y_2 Y_3  \\
	Y_3^2 -Y_1 Y_2 
	\end{pmatrix}.
	\end{aligned}
	\label{weight8}
	\end{align}
	For weigh 10, there are  eleven modular forms as:
	\begin{align}
	&{\bf Y^{(\rm 10)}_1}=(Y_1^2+2Y_2 Y_3) (Y_1^3+ Y_2^3+Y_3^3 -3Y_1 Y_2 Y_3) \,\nonumber\\
	&{\bf Y^{(\rm 10)}_{1'}}=(Y_3^2+2Y_1 Y_2) (Y_1^3+ Y_2^3+Y_3^3 -3Y_1 Y_2 Y_3)\,,
	\nonumber \\
	& {\bf Y^{(\rm 10)}_{3,{\rm 1}}}\equiv 
	\begin{pmatrix}
	Y_{1,1}^{(10)}  \\
	Y_{2,1}^{(10)} \\
	Y_{3,1}^{(10)}
	\end{pmatrix} 
	=(Y_1^2+2 Y_2 Y_3)^2 
	\begin{pmatrix}
	Y_1  \\
	Y_2 \\
	Y_3
	\end{pmatrix} \,,\nonumber\\
	& {\bf Y^{(\rm 10)}_{3,{\rm 2}}}\equiv 
	\begin{pmatrix}
	Y_{1,2}^{(10)}  \\
	Y_{2,2}^{(10)} \\
	Y_{3,2}^{(10)}
	\end{pmatrix} 
	=(Y_3^2+2 Y_1 Y_2 )^2 
	\begin{pmatrix}
	Y_2 \\
	Y_3  \\
	Y_1 
	\end{pmatrix}\,,\nonumber\\
	& {\bf Y^{(\rm 10)}_{3,{\rm 3}}}\equiv 
	\begin{pmatrix}
	Y_{1,3}^{(10)}  \\
	Y_{2,3}^{(10)} \\
	Y_{3,3}^{(10)}
	\end{pmatrix} 
	=(Y_1^2+2 Y_2 Y_3 )(Y_3^2+2 Y_1 Y_2 )
	\begin{pmatrix}
	Y_3 \\
	Y_1  \\
	Y_2 
	\end{pmatrix}\,.
	\label{weight10}
	\end{align}
	%
	
	At the fixed point  $\tau=\omega$, they are given as:
	\begin{align}
	&\begin{aligned}
	{\bf Y^{(\rm 2)}_3}
	=Y_0
	\begin{pmatrix}1\\ \omega\\ -\frac12 \omega^2\end{pmatrix}\,,
	\end{aligned} \quad 
	\begin{aligned} {\bf Y^{(\rm 4)}_{3}}=
	\frac32 Y_0^2 
	\begin{pmatrix}
	1  \\
	-\frac12 \omega \\
	\omega^2
	\end{pmatrix}\ , \quad {\bf Y^{(\rm 4)}_1}=0 \ , \quad
	{\bf Y^{(\rm 4)}_{1'}}=\frac94 Y_0^2 \, \omega \ , 
	\end{aligned} \nonumber\\
	&
	\begin{aligned}
	{\bf Y^{(\rm 6)}_{3}}=0
	\end{aligned}\ , \quad
	\begin{aligned} {\bf Y^{(\rm 6)}_{3'}}=
	\frac98Y_0^3 
	\begin{pmatrix}
	-1  \\
	2\omega \\
	2\omega^2
	\end{pmatrix}\ , \quad{\bf Y^{(\rm 6)}_1}=\frac{27}{8} Y_0^3 \, , 
	\end{aligned}\nonumber\\
	&\begin{aligned}
	{\bf Y^{(\rm 8)}_{3}}=0
	\end{aligned}\ , \quad 
	\begin{aligned} {\bf Y^{(\rm 8)}_{3'}}=
	\frac{27}{8}Y_0^4 
	\begin{pmatrix}
	1  \\
	\omega \\
	-\frac12\omega^2
	\end{pmatrix}\ , \quad {\bf Y^{(\rm 8)}_1}=0, \quad {\bf Y^{(\rm 8)}_{1'}}=0 \, , 
	\quad {\bf Y^{(\rm 8)}_{1''}}=\frac94\,\omega Y_0^4 \, ,
	\end{aligned}\nonumber\\
	&\begin{aligned}
	{\bf Y^{(\rm 10)}_{3,{\rm 1}}}=0
	\end{aligned} , \  
	\begin{aligned} {\bf Y^{(\rm 10)}_{3,{\rm 2}}}=
	\frac{81}{16}\omega^2 Y_0^5 
	\begin{pmatrix}
	\omega \\
	-\frac12\omega^2\\
	1  
	\end{pmatrix} ,
	\end{aligned} \ 
	\begin{aligned}
	{\bf Y^{(\rm 10)}_{3,{\rm 3}}}=0, \ 
	{\bf Y^{(\rm 10}_1}=0, 
	\  {\bf Y^{(\rm 10)}_{1'}}=\frac{243}{32}\,\omega Y_0^5 .
	\end{aligned}
	\label{tauomega}
	\end{align}

\section{Tensor product of  $\rm A_4$ group}
	%
	%
	\label{Tensor}
	
	
	We take the generators of $A_4$ group for the triplet as follows:
	\begin{align}
	\begin{aligned}
	S=\frac{1}{3}
	\begin{pmatrix}
	-1 & 2 & 2 \\
	2 &-1 & 2 \\
	2 & 2 &-1
	\end{pmatrix},
	\end{aligned}
	\qquad 
	\begin{aligned}
	T=
	\begin{pmatrix}
	1 & 0& 0 \\
	0 &\omega& 0 \\
	0 & 0 & \omega^2
	\end{pmatrix}, 
	\end{aligned}
	\label{ST}
	\end{align}
	where $\omega=e^{i\frac{2}{3}\pi}$ for a triplet.
	In this base,
	the multiplication rule is
	\begin{align}
	\begin{pmatrix}
	a_1\\
	a_2\\
	a_3
	\end{pmatrix}_{\bf 3}
	\otimes 
	\begin{pmatrix}
	b_1\\
	b_2\\
	b_3
	\end{pmatrix}_{\bf 3}
	&=\left (a_1b_1+a_2b_3+a_3b_2\right )_{\bf 1} 
	\oplus \left (a_3b_3+a_1b_2+a_2b_1\right )_{{\bf 1}'} \nonumber \\
	& \oplus \left (a_2b_2+a_1b_3+a_3b_1\right )_{{\bf 1}''} \nonumber \\
	&\oplus \frac13
	\begin{pmatrix}
	2a_1b_1-a_2b_3-a_3b_2 \\
	2a_3b_3-a_1b_2-a_2b_1 \\
	2a_2b_2-a_1b_3-a_3b_1
	\end{pmatrix}_{{\bf 3}}
	\oplus \frac12
	\begin{pmatrix}
	a_2b_3-a_3b_2 \\
	a_1b_2-a_2b_1 \\
	a_3b_1-a_1b_3
	\end{pmatrix}_{{\bf 3}\  } \ , \nonumber \\
	\nonumber \\
	{\bf 1} \otimes {\bf 1} = {\bf 1} \ , \qquad &
	{\bf 1'} \otimes {\bf 1'} = {\bf 1''} \ , \qquad
	{\bf 1''} \otimes {\bf 1''} = {\bf 1'} \ , \qquad
	{\bf 1'} \otimes {\bf 1''} = {\bf 1} \  ,
	\end{align}
	where
	\begin{align}
	T({\bf 1')}=\omega\,,\qquad T({\bf 1''})=\omega^2. 
	\end{align}
	More details are shown in the review~
	\cite{Ishimori:2010au,Ishimori:2012zz,Kobayashi:2022moq}.
	
\newpage


\begin{thebibliography}{99}
	
	\bibitem{Weinberg:1977hb}
	S.~Weinberg,
	Trans. New York Acad. Sci. \textbf{38} (1977), 185-201.
	
	\bibitem{Fritzsch:1977za}
	H.~Fritzsch,
	Phys. Lett. B \textbf{70} (1977), 436-440.
	
	
	
	
	
	
	
	
	\bibitem{Fritzsch:1977vd}
	H.~Fritzsch,
	Phys. Lett. B \textbf{73} (1978), 317-322.
	
	
	\bibitem{Fritzsch:1979zq}
	H.~Fritzsch,
	Nucl. Phys. B \textbf{155} (1979), 189-207.
	
	
	
	\bibitem{Georgi:1979df}
	H.~Georgi and C.~Jarlskog,
	Phys. Lett. B \textbf{86} (1979), 297-300.
	
	\bibitem{Branco:1988iq}
	G.~C.~Branco, L.~Lavoura and F.~Mota,
	Phys. Rev. D \textbf{39} (1989), 3443.
	
	\bibitem{Dimopoulos:1991za}
	S.~Dimopoulos, L.~J.~Hall and S.~Raby,
	Phys. Rev. D \textbf{45} (1992), 4192-4200.
	
	\bibitem{Ramond:1993kv}
	P.~Ramond, R.~G.~Roberts and G.~G.~Ross,
	Nucl. Phys. B \textbf{406} (1993), 19-42.
	[arXiv:hep-ph/9303320 [hep-ph]].
	
	\bibitem{Frampton:2002yf}
	P.~H.~Frampton, S.~L.~Glashow and D.~Marfatia,
	Phys. Lett. B \textbf{536} (2002), 79-82
	[arXiv:hep-ph/0201008 [hep-ph]].
	
	\bibitem{Froggatt:1978nt}
	C.~D.~Froggatt and H.~B.~Nielsen,
	Nucl. Phys. B \textbf{147} (1979), 277-298
	
	
	\bibitem{Pakvasa:1977in}
	S.~Pakvasa and H.~Sugawara,
	Phys.\ Lett.\  {\bf 73B} (1978) 61.
	
	\bibitem{Wilczek:1977uh}
	F.~Wilczek and A.~Zee,
	Phys.\ Lett.\  {\bf 70B} (1977) 418
	Erratum: [Phys.\ Lett.\  {\bf 72B} (1978) 504].
	
	\bibitem{Fukugita:1998vn}
	M.~Fukugita, M.~Tanimoto and T.~Yanagida,
	Phys.\ Rev.\ D {\bf 57} (1998) 4429
	[hep-ph/9709388].
	
	\bibitem{Fukuda:1998mi}
	Y.~Fukuda {\it et al.} [Super-Kamiokande Collaboration],
	Phys.\ Rev.\ Lett.\  {\bf 81} (1998) 1562
	[hep-ex/9807003].
	
	
	\bibitem{Altarelli:2010gt}
	G.~Altarelli and F.~Feruglio,
	Rev.\ Mod.\ Phys.\  {\bf 82} (2010) 2701
	[arXiv:1002.0211 [hep-ph]].
	
	
	
	\bibitem{Ishimori:2010au}
	H.~Ishimori, T.~Kobayashi, H.~Ohki, Y.~Shimizu, H.~Okada and M.~Tanimoto,
	Prog.\ Theor.\ Phys.\ Suppl.\  {\bf 183} (2010) 1
	[arXiv:1003.3552 [hep-th]].
	
	
	
	\bibitem{Ishimori:2012zz}
	H.~Ishimori, T.~Kobayashi, H.~Ohki, H.~Okada, Y.~Shimizu and M.~Tanimoto,
	Lect.\ Notes Phys.\  {\bf 858} (2012) 1, Springer.
	
	\bibitem{Kobayashi:2022moq}
	T.~Kobayashi, H.~Ohki, H.~Okada, Y.~Shimizu and M.~Tanimoto,
	Lect.\ Notes Phys.\ {\bf 995} (2022) 1, Springer 
	doi:10.1007/978-3-662-64679-3.
	
	\bibitem{Hernandez:2012ra}
	D.~Hernandez and A.~Y.~Smirnov,
	Phys.\ Rev.\ D {\bf 86} (2012) 053014
	[arXiv:1204.0445 [hep-ph]].
	
	\bibitem{King:2013eh}
	S.~F.~King and C.~Luhn,
	Rept.\ Prog.\ Phys.\  {\bf 76} (2013) 056201
	[arXiv:1301.1340 [hep-ph]].
	
	\bibitem{King:2014nza} 
	S.~F.~King, A.~Merle, S.~Morisi, Y.~Shimizu and M.~Tanimoto,
	New J.\ Phys.\  {\bf 16}, 045018 (2014)
	[arXiv:1402.4271 [hep-ph]].
	
	
	\bibitem{Tanimoto:2015nfa}
	M.~Tanimoto,
	AIP Conf.\ Proc.\  {\bf 1666} (2015) 120002.
	
	\bibitem{King:2017guk}
	S.~F.~King,
	Prog.\ Part.\ Nucl.\ Phys.\  {\bf 94} (2017) 217
	[arXiv:1701.04413 [hep-ph]].
	
	\bibitem{Petcov:2017ggy}
	S.~T.~Petcov,
	Eur.\ Phys.\ J.\ C {\bf 78} (2018) no.9,  709
	[arXiv:1711.10806 [hep-ph]].
	
	\bibitem{Feruglio:2019ktm}
	F.~Feruglio and A.~Romanino,
	arXiv:1912.06028 [hep-ph].
	
	
	
	
	
	\bibitem{Buchmuller:1985jz}
	W.~Buchmuller and D.~Wyler,
	Nucl. Phys. B \textbf{268} (1986), 621-653.
	
	\bibitem{Grzadkowski:2010es}
	B.~Grzadkowski, M.~Iskrzynski, M.~Misiak and J.~Rosiek,
	JHEP \textbf{10} (2010), 085
	[arXiv:1008.4884 [hep-ph]].
	
	
	\bibitem{Alonso:2013hga}
	R.~Alonso, E.~E.~Jenkins, A.~V.~Manohar and M.~Trott,
	JHEP \textbf{04} (2014), 159
	[arXiv:1312.2014 [hep-ph]].
	
	
	\bibitem{Faroughy:2020ina}
	D.~A.~Faroughy, G.~Isidori, F.~Wilsch and K.~Yamamoto,
	JHEP \textbf{08} (2020), 166
	[arXiv:2005.05366 [hep-ph]].
	
	\bibitem{Gerard:1982mm}
	J.~M.~Gerard,
	Z. Phys. C \textbf{18} (1983), 145.
	
	
	\bibitem{Chivukula:1987py}
	R.~S.~Chivukula and H.~Georgi,
	Phys. Lett. B \textbf{188} (1987), 99-104.
	
	\bibitem{DAmbrosio:2002vsn}
	G.~D'Ambrosio, G.~F.~Giudice, G.~Isidori and A.~Strumia,
	Nucl. Phys. B \textbf{645} (2002), 155-187
	[arXiv:hep-ph/0207036 [hep-ph]].
	
	\bibitem{Barbieri:2011ci}
	R.~Barbieri, G.~Isidori, J.~Jones-Perez, P.~Lodone and D.~M.~Straub,
	Eur. Phys. J. C \textbf{71} (2011), 1725
	[arXiv:1105.2296 [hep-ph]].
	
	\bibitem{Barbieri:2012uh}
	R.~Barbieri, D.~Buttazzo, F.~Sala and D.~M.~Straub,
	JHEP \textbf{07} (2012), 181
	[arXiv:1203.4218 [hep-ph]].
	
	\bibitem{Blankenburg:2012nx}
	G.~Blankenburg, G.~Isidori and J.~Jones-Perez,
	Eur. Phys. J. C \textbf{72} (2012), 2126
	[arXiv:1204.0688 [hep-ph]].
	
	
	
	
	\bibitem{Kobayashi:2004ya}
	T.~Kobayashi, S.~Raby and R.~J.~Zhang,
	Nucl. Phys. B \textbf{704}, 3-55 (2005)
	[arXiv:hep-ph/0409098 [hep-ph]].
	
	\bibitem{Kobayashi:2006wq}
	T.~Kobayashi, H.~P.~Nilles, F.~Ploger, S.~Raby and M.~Ratz,
	Nucl. Phys. B \textbf{768}, 135-156 (2007)
	[arXiv:hep-ph/0611020 [hep-ph]].
	
	\bibitem{Ko:2007dz}
	P.~Ko, T.~Kobayashi, J.~h.~Park and S.~Raby,
	Phys. Rev. D \textbf{76}, 035005 (2007)
	[erratum: Phys. Rev. D \textbf{76}, 059901 (2007)]
	[arXiv:0704.2807 [hep-ph]].
	
	\bibitem{Abe:2009vi}
	H.~Abe, K.~S.~Choi, T.~Kobayashi and H.~Ohki,
	Nucl. Phys. B \textbf{820}, 317-333 (2009)
	[arXiv:0904.2631 [hep-ph]].
	
	\bibitem{Beye:2014nxa}
	F.~Beye, T.~Kobayashi and S.~Kuwakino,
	Phys. Lett. B \textbf{736}, 433-437 (2014)
	[arXiv:1406.4660 [hep-th]].
	
	
	
	
	\bibitem{Ferrara:1989qb}
	S.~Ferrara, D.~Lust and S.~Theisen,
	Phys. Lett. B \textbf{233} (1989), 147-152.
	
	\bibitem{Lerche:1989cs}
	W.~Lerche, D.~Lust and N.~P.~Warner,
	Phys. Lett. B \textbf{231} (1989), 417-424.
	%
	
	\bibitem{Lauer:1989ax}
	J.~Lauer, J.~Mas and H.~P.~Nilles,
	Phys. Lett. B \textbf{226}, 251-256 (1989)
	doi:10.1016/0370-2693(89)91190-8.
	
	
	\bibitem{Lauer:1990tm} 
	J.~Lauer, J.~Mas and H.~P.~Nilles,
	Nucl.\ Phys.\ B {\bf 351}, 353 (1991).
	
	
	
	
	
	
	\bibitem{Kobayashi:2018rad} 
	T.~Kobayashi, S.~Nagamoto, S.~Takada, S.~Tamba and T.~H.~Tatsuishi,
	Phys.\ Rev.\ D {\bf 97}, no. 11, 116002 (2018)
	[arXiv:1804.06644 [hep-th]].
	
	\bibitem{Kobayashi:2018bff}
	T.~Kobayashi and S.~Tamba,
	Phys. Rev. D \textbf{99}, no.4, 046001 (2019)
	[arXiv:1811.11384 [hep-th]].
	
	\bibitem{Ohki:2020bpo}
	H.~Ohki, S.~Uemura and R.~Watanabe,
	Phys. Rev. D \textbf{102}, no.8, 085008 (2020)
	[arXiv:2003.04174 [hep-th]].
	
	
	
	\bibitem{Kikuchi:2020frp}
	S.~Kikuchi, T.~Kobayashi, S.~Takada, T.~H.~Tatsuishi and H.~Uchida,
	Phys. Rev. D \textbf{102}, no.10, 105010 (2020)
	[arXiv:2005.12642 [hep-th]].
	
	
	\bibitem{Kikuchi:2020nxn}
	S.~Kikuchi, T.~Kobayashi, H.~Otsuka, S.~Takada and H.~Uchida,
	JHEP \textbf{11}, 101 (2020)
	[arXiv:2007.06188 [hep-th]].
	
	
	\bibitem{Kikuchi:2021ogn}
	S.~Kikuchi, T.~Kobayashi and H.~Uchida,
	Phys. Rev. D \textbf{104}, no.6, 065008 (2021)
	[arXiv:2101.00826 [hep-th]].
	
	
	
	\bibitem{Almumin:2021fbk}
	Y.~Almumin, M.~C.~Chen, V.~Knapp-Perez, S.~Ramos-Sanchez, M.~Ratz and S.~Shukla,
	JHEP \textbf{05} (2021), 078
	[arXiv:2102.11286 [hep-th]].
	
	
	
	
	\bibitem{Strominger:1990pd}
	A.~Strominger,
	Commun. Math. Phys. \textbf{133} (1990), 163-180.
	
	\bibitem{Candelas:1990pi}
	P.~Candelas and X.~de la Ossa,
	Nucl. Phys. B \textbf{355} (1991), 455-481.
	
	\bibitem{Ishiguro:2020nuf}
	K.~Ishiguro, T.~Kobayashi and H.~Otsuka,
	Nucl. Phys. B \textbf{973}, 115598 (2021)
	[arXiv:2010.10782 [hep-th]].
	
	
	\bibitem{Ishiguro:2021ccl}
	K.~Ishiguro, T.~Kobayashi and H.~Otsuka,
	JHEP \textbf{01}, 020 (2022)
	[arXiv:2107.00487 [hep-th]].
	
	
	
	\bibitem{deAdelhartToorop:2011re} 
	R.~de Adelhart Toorop, F.~Feruglio and C.~Hagedorn,
	Nucl.\ Phys.\ B {\bf 858}, 437 (2012)
	[arXiv:1112.1340 [hep-ph]].
	
	\bibitem{Feruglio:2017spp}
	F.~Feruglio,
	in From My Vast Repertoire ...: Guido Altarelli's Legacy,
	A. Levy,  S. Forte, Stefano, and G. Ridolfi, eds.,
	pp.227--266, 2019, arXiv:1706.08749 [hep-ph]. 
	
	
	
	\bibitem{Kobayashi:2018vbk}
	T.~Kobayashi, K.~Tanaka and T.~H.~Tatsuishi,
	Phys.\ Rev.\ D {\bf 98} (2018) no.1,  016004
	[arXiv:1803.10391 [hep-ph]].
	
	\bibitem{Penedo:2018nmg}
	J.~T.~Penedo and S.~T.~Petcov,
	Nucl.\ Phys.\ B {\bf 939} (2019) 292
	[arXiv:1806.11040 [hep-ph]].
	
	\bibitem{Novichkov:2018nkm} 
	P.~P.~Novichkov, J.~T.~Penedo, S.~T.~Petcov and A.~V.~Titov,
	JHEP {\bf 1904}, 174 (2019)
	[arXiv:1812.02158 [hep-ph]].
	
	
	\bibitem{Ding:2019xna}
	G.~J.~Ding, S.~F.~King and X.~G.~Liu,
	Phys.\ Rev.\ D {\bf 100} (2019) no.11,  115005
	[arXiv:1903.12588 [hep-ph]].
	
	
	
	\bibitem{Green:1987mn} 
	M.~B.~Green, J.~H.~Schwarz and E.~Witten,
	Cambridge, Uk: Univ. Pr. ( 1987) 596 P. ( Cambridge Monographs On Mathematical Physics)
	
	\bibitem{Strominger:1985it} 
	A.~Strominger and E.~Witten,
	Commun.\ Math.\ Phys.\ {\bf 101}, 341 (1985).
	
	
	\bibitem{Dine:1992ya} 
	M.~Dine, R.~G.~Leigh and D.~A.~MacIntire,
	Phys.\ Rev.\ Lett.\ {\bf 69}, 2030 (1992)
	[hep-th/9205011].
	
	\bibitem{Choi:1992xp} 
	K.~w.~Choi, D.~B.~Kaplan and A.~E.~Nelson,
	Nucl.\ Phys.\ B {\bf 391}, 515 (1993)
	[hep-ph/9205202].
	
	\bibitem{Lim:1990bp} 
	C.~S.~Lim,
	Phys.\ Lett.\ B {\bf 256}, 233 (1991).
	
	\bibitem{Kobayashi:1994ks} 
	T.~Kobayashi and C.~S.~Lim,
	Phys.\ Lett.\ B {\bf 343}, 122 (1995)
	[hep-th/9410023].
	
	
	
	\bibitem{Acharya:1995ag} 
	B.~S.~Acharya, D.~Bailin, A.~Love, W.~A.~Sabra and S.~Thomas,
	Phys.\ Lett.\ B {\bf 357}, 387 (1995)
	Erratum: [Phys.\ Lett.\ B {\bf 407}, 451 (1997)]
	[hep-th/9506143].
	
	
	\bibitem{Dent:2001cc} 
	T.~Dent,
	Phys.\ Rev.\ D {\bf 64}, 056005 (2001)
	[hep-ph/0105285].
	
	\bibitem{Khalil:2001dr} 
	S.~Khalil, O.~Lebedev and S.~Morris,
	Phys.\ Rev.\ D {\bf 65}, 115014 (2002)
	[hep-th/0110063].
	
	
	\bibitem{Giedt:2002ns} 
	J.~Giedt,
	Mod.\ Phys.\ Lett.\ A {\bf 17}, 1465 (2002)
	[hep-ph/0204017].
	
	
	
	
	\bibitem{Kobayashi:2020uaj}
	T.~Kobayashi and H.~Otsuka,
	Phys. Rev. D \textbf{102}, no.2, 026004 (2020)
	[arXiv:2004.04518 [hep-th]].
	
	
	
	
	
	\bibitem{Baur:2019kwi} 
	A.~Baur, H.~P.~Nilles, A.~Trautner and P.~K.~S.~Vaudrevange,
	Phys.\ Lett.\ B {\bf 795}, 7 (2019)
	[arXiv:1901.03251 [hep-th]]; 
	
	
	\bibitem{Novichkov:2019sqv} 
	P.~P.~Novichkov, J.~T.~Penedo, S.~T.~Petcov and A.~V.~Titov,
	JHEP {\bf 1907}, 165 (2019)
	[arXiv:1905.11970 [hep-ph]].
	
	
	\bibitem{Baur:2019iai}
	A.~Baur, H.~P.~Nilles, A.~Trautner and P.~K.~S.~Vaudrevange,
	Nucl. Phys. B \textbf{947} (2019), 114737
	[arXiv:1908.00805 [hep-th]].
	
	
	
	\bibitem{Baur:2020jwc}
	A.~Baur, M.~Kade, H.~P.~Nilles, S.~Ramos-Sanchez and P.~K.~S.~Vaudrevange,
	JHEP \textbf{02} (2021), 018
	[arXiv:2008.07534 [hep-th]].
	
	\bibitem{Nilles:2020gvu}
	H.~P.~Nilles, S.~Ramos\textendash{}S\'anchez and P.~K.~S.~Vaudrevange,
	Nucl. Phys. B \textbf{966} (2021), 115367
	[arXiv:2010.13798 [hep-th]].
	
	
	
	
	
	
	
	
	\bibitem{Bonisch:2022slo}
	K.~B\"onisch, M.~Elmi, A.~K.~Kashani-Poor and A.~Klemm,
	[arXiv:2204.06506 [hep-th]].
	
	
	
	\bibitem{Kobayashi:2019uyt}
	T.~Kobayashi, Y.~Shimizu, K.~Takagi, M.~Tanimoto, T.~H.~Tatsuishi and H.~Uchida,
	Phys. Rev. D \textbf{101}, no.5, 055046 (2020)
	[arXiv:1910.11553 [hep-ph]].
	
	
	
	\bibitem{Kikuchi:2022geu}
	S.~Kikuchi, T.~Kobayashi, M.~Tanimoto and H.~Uchida,
	PTEP \textbf{2022}, no.11, 113B07 (2022)
	[arXiv:2206.08538 [hep-ph]].
	
	
	
	\bibitem{Kobayashi:2021uam}
	T.~Kobayashi and H.~Otsuka,
	Eur. Phys. J. C \textbf{82}, no.1, 25 (2022)
	[arXiv:2108.02700 [hep-ph]].
	
	
	\bibitem{Kobayashi:2021pav}
	T.~Kobayashi, H.~Otsuka, M.~Tanimoto and K.~Yamamoto,
	Phys. Rev. D \textbf{105}, no.5, 055022 (2022)
	[arXiv:2112.00493 [hep-ph]].
	
	
	\bibitem{Kobayashi:2022jvy}
	T.~Kobayashi, H.~Otsuka, M.~Tanimoto and K.~Yamamoto,
	JHEP \textbf{08}, 013 (2022)
	[arXiv:2204.12325 [hep-ph]].
	
	
	
	
	\bibitem{Nomura:2019jxj}
	T.~Nomura and H.~Okada,
	Phys. Lett. B \textbf{797}, 134799 (2019)
	[arXiv:1904.03937 [hep-ph]].
	
	
	\bibitem{Kobayashi:2021ajl}
	T.~Kobayashi, H.~Okada and Y.~Orikasa,
	Phys. Dark Univ. \textbf{37}, 101080 (2022)
	[arXiv:2111.05674 [hep-ph]].
	
	
	
	\bibitem{Kobayashi:2021jqu}
	T.~Kobayashi, T.~Shimomura and M.~Tanimoto,
	Phys. Lett. B \textbf{819}, 136452 (2021)
	[arXiv:2102.10425 [hep-ph]].
	
	
	\bibitem{Tanimoto:2021ehw}
	M.~Tanimoto and K.~Yamamoto,
	JHEP \textbf{10}, 183 (2021)
	[arXiv:2106.10919 [hep-ph]].
	
	
	\bibitem{Kikuchi:2022pkd}
	S.~Kikuchi, T.~Kobayashi, K.~Nasu, H.~Otsuka, S.~Takada and H.~Uchida,
	PTEP \textbf{2022}, no.12, 123B02 (2022)
	[arXiv:2203.14667 [hep-ph]].
	
	
	\bibitem{Kobayashi:2022sov}
	T.~Kobayashi, S.~Nishimura, H.~Otsuka, M.~Tanimoto and K.~Yamamoto,
	[arXiv:2207.14014 [hep-ph]].
	
	\bibitem{Feruglio:2023uof}
	F.~Feruglio, A.~Strumia and A.~Titov,
	[arXiv:2305.08908 [hep-ph]].
	
	
	
	
	\bibitem{Gunning:1962}
	R. C. Gunning,
	\textit{Lectures on Modular Forms}
	(Princeton University Press, Princeton, NJ, 1962).
	
	
	\bibitem{Schoeneberg:1974}
	B.~Schoeneberg,
	\textit{Elliptic Modular Functions}
	(Springer-Verlag, 1974).
	
	\bibitem{Koblitz:1984}
	N.~Koblitz,
	\textit{Introduction to Elliptic Curves and Modular Forms}
	(Springer-Verlag, 1984).
	
	\bibitem{Bruinier:2008}
	J.H.~Bruinier, G.V.D.~Geer, G.~Harder, and D.~Zagier,
	\textit{The 1-2-3 of Modular Forms}
	(Springer, 2008).
	
	
	
	\bibitem{Ferrara:1989bc} 
	S.~Ferrara, D.~Lust, A.~D.~Shapere and S.~Theisen,
	Phys.\ Lett.\ B {\bf 225}, 363 (1989).
	
	\bibitem{Chen:2019ewa}
	M.~Chen, S.~Ramos-S\'anchez and M.~Ratz,
	Phys. Lett. B \textbf{801} (2020), 135153
	[arXiv:1909.06910 [hep-ph]].
	
	
	
	
	\bibitem{Kobayashi:2018scp}
	T.~Kobayashi, N.~Omoto, Y.~Shimizu, K.~Takagi, M.~Tanimoto and T.~H.~Tatsuishi,
	JHEP \textbf{11} (2018), 196
	[arXiv:1808.03012 [hep-ph]].
	
	
	\bibitem{Esteban:2020cvm}
	I.~Esteban, M.~C.~Gonzalez-Garcia, M.~Maltoni, T.~Schwetz and A.~Zhou,
	JHEP \textbf{09} (2020), 178
	[arXiv:2007.14792 [hep-ph]].
	
	\bibitem{Vagnozzi:2017ovm}
	S.~Vagnozzi, E.~Giusarma, O.~Mena, K.~Freese, M.~Gerbino, S.~Ho and M.~Lattanzi,
	Phys.\ Rev.\ D {\bf 96} (2017) no.12,  123503
	[arXiv:1701.08172 [astro-ph.CO]].
	
	
	\bibitem{Aghanim:2018eyx}
	N.~Aghanim \textit{et al.} [Planck],
	Astron. Astrophys. \textbf{641} (2020), A6
	[arXiv:1807.06209 [astro-ph.CO]].
	
	\bibitem{ParticleDataGroup:2022pth}
	R.~L.~Workman \textit{et al.} [Particle Data Group],
	PTEP \textbf{2022} (2022), 083C01.
	
	
	\bibitem{Antusch:2013jca}
	S.~Antusch and V.~Maurer,
	JHEP {\bf 1311} (2013) 115
	[arXiv:1306.6879 [hep-ph]].
	
	\bibitem{Bjorkeroth:2015ora}
	F.~Bj\"orkeroth, F.~J.~de Anda, I.~de Medeiros Varzielas and S.~F.~King,
	JHEP {\bf 1506} (2015) 141
	[arXiv:1503.03306 [hep-ph]].
	
	\bibitem{Criado:2018thu}
	J.~C.~Criado and F.~Feruglio,
	SciPost Phys.\  {\bf 5} (2018) no.5,  042
	[arXiv:1807.01125 [hep-ph]].
	
	\bibitem{Ding:2019zxk}
	G.~J.~Ding, S.~F.~King and X.~G.~Liu,
	JHEP {\bf 1909} (2019) 074
	[arXiv:1907.11714 [hep-ph]].
	
	\bibitem{Okada:2020brs}
	H.~Okada and M.~Tanimoto,
	JHEP \textbf{03} (2021), 010
	[arXiv:2012.01688 [hep-ph]].
	
	
	\bibitem{Yao:2020qyy}
	C.~Y.~Yao, J.~N.~Lu and G.~J.~Ding,
	JHEP \textbf{05} (2021), 102
	[arXiv:2012.13390 [hep-ph]].
	
	
	
	\bibitem{Novichkov:2018ovf}
	P.~P.~Novichkov, J.~T.~Penedo, S.~T.~Petcov and A.~V.~Titov,
	JHEP {\bf 1904} (2019) 005
	[arXiv:1811.04933 [hep-ph]].
	
	\bibitem{Kobayashi:2019mna}
	T.~Kobayashi, Y.~Shimizu, K.~Takagi, M.~Tanimoto and T.~H.~Tatsuishi,
	JHEP \textbf{02} (2020), 097
	[arXiv:1907.09141 [hep-ph]].
	
	
	
	\bibitem{Wang:2019ovr}
	X.~Wang and S.~Zhou,
	JHEP \textbf{05} (2020), 017
	[arXiv:1910.09473 [hep-ph]].
	
	
	
	\bibitem{Gehrlein:2020jnr}
	J.~Gehrlein and M.~Spinrath,
	JHEP \textbf{03} (2021), 177
	[arXiv:2012.04131 [hep-ph]].
	
	
	\bibitem{Kobayashi:2019rzp}
	T.~Kobayashi, Y.~Shimizu, K.~Takagi, M.~Tanimoto and T.~H.~Tatsuishi,
	PTEP \textbf{2020}, no.5, 053B05 (2020)
	[arXiv:1906.10341 [hep-ph]].
	
	\bibitem{Meloni:2023aru}
	D.~Meloni and M.~Parriciatu,
	[arXiv:2306.09028 [hep-ph]].
	
	
	
	\bibitem{Liu:2019khw}
	X.~G.~Liu and G.~J.~Ding,
	JHEP {\bf 1908} (2019) 134
	[arXiv:1907.01488 [hep-ph]].
	
	
	
	
	
	
	\bibitem{Novichkov:2020eep}
	P.~P.~Novichkov, J.~T.~Penedo and S.~T.~Petcov,
	Nucl. Phys. B \textbf{963} (2021), 115301
	[arXiv:2006.03058 [hep-ph]].
	
	
	
	\bibitem{Liu:2020akv}
	X.~G.~Liu, C.~Y.~Yao and G.~J.~Ding,
	Phys. Rev. D \textbf{103} (2021) no.5, 056013
	[arXiv:2006.10722 [hep-ph]].
	
	\bibitem{Wang:2020lxk}
	X.~Wang, B.~Yu and S.~Zhou,
	Phys. Rev. D \textbf{103} (2021) no.7, 076005
	[arXiv:2010.10159 [hep-ph]].
	
	
	
	
	\bibitem{Yao:2020zml}
	C.~Y.~Yao, X.~G.~Liu and G.~J.~Ding,
	[arXiv:2011.03501 [hep-ph]].
	
	
	\bibitem{Okada:2022kee}
	H.~Okada and Y.~Orikasa,
	Chin. Phys. C \textbf{46} (2022) no.12, 123108
	[arXiv:2206.12629 [hep-ph]].
	
	
	\bibitem{Ding:2022nzn}
	G.~J.~Ding, X.~G.~Liu and C.~Y.~Yao,
	[arXiv:2211.04546 [hep-ph]].
	
	\bibitem{Ding:2022aoe}
	G.~J.~Ding, F.~R.~Joaquim and J.~N.~Lu,
	[arXiv:2211.08136 [hep-ph]].
	
	
	
	\bibitem{Benes:2022bbg}
	P.~Bene\v{s}, H.~Okada and Y.~Orikasa,
	[arXiv:2212.07245 [hep-ph]].
	
	\bibitem{Ding:2020msi}
	G.~J.~Ding, S.~F.~King, C.~C.~Li and Y.~L.~Zhou,
	JHEP \textbf{08} (2020), 164
	[arXiv:2004.12662 [hep-ph]].
	
	\bibitem{Li:2021buv}
	C.~C.~Li, X.~G.~Liu and G.~J.~Ding,
	JHEP \textbf{10} (2021), 238
	[arXiv:2108.02181 [hep-ph]].
	
	
	
	
	\bibitem{Okada:2018yrn}
	H.~Okada and M.~Tanimoto,
	Phys.\ Lett.\ B {\bf 791} (2019) 54
	[arXiv:1812.09677 [hep-ph]].
	
	\bibitem{Okada:2019uoy}
	H.~Okada and M.~Tanimoto,
	Eur. Phys. J. C \textbf{81} (2021) no.1, 52
	[arXiv:1905.13421 [hep-ph]].
	
	
	
	\bibitem{deAnda:2018ecu}
	F.~J.~de Anda, S.~F.~King and E.~Perdomo,
	Phys. Rev. D \textbf{101} (2020) no.1, 015028
	[arXiv:1812.05620 [hep-ph]].
	
	
	
	\bibitem{King:2021fhl}
	S.~F.~King and Y.~L.~Zhou,
	JHEP \textbf{04} (2021), 291
	[arXiv:2103.02633 [hep-ph]].
	
	\bibitem{Chen:2021zty}
	P.~Chen, G.~J.~Ding and S.~F.~King,
	JHEP \textbf{04} (2021), 239
	[arXiv:2101.12724 [hep-ph]].
	
	
	\bibitem{Du:2020ylx}
	X.~Du and F.~Wang,
	JHEP \textbf{02}, 221 (2021)
	[arXiv:2012.01397 [hep-ph]].
	
	
	
	\bibitem{Ding:2021zbg}
	G.~J.~Ding, S.~F.~King and C.~Y.~Yao,
	[arXiv:2103.16311 [hep-ph]].
	
	\bibitem{Ding:2021eva}
	G.~J.~Ding, S.~F.~King and J.~N.~Lu,
	JHEP \textbf{11} (2021), 007
	[arXiv:2108.09655 [hep-ph]].
	
	
	\bibitem{Ding:2022bzs}
	G.~J.~Ding, S.~F.~King, J.~N.~Lu and B.~Y.~Qu,
	JHEP \textbf{10} (2022), 071
	[arXiv:2206.14675 [hep-ph]].
	
	
	\bibitem{Asaka:2019vev}
	T.~Asaka, Y.~Heo, T.~H.~Tatsuishi and T.~Yoshida,
	JHEP {\bf 2001} (2020) 144
	[arXiv:1909.06520 [hep-ph]].
	
	
	\bibitem{Okada:2021qdf}
	H.~Okada, Y.~Shimizu, M.~Tanimoto and T.~Yoshida,
	JHEP \textbf{07} (2021), 184
	[arXiv:2105.14292 [hep-ph]].
	
	\bibitem{Qu:2021jdy}
	B.~Y.~Qu, X.~G.~Liu, P.~T.~Chen and G.~J.~Ding,
	Phys. Rev. D \textbf{104} (2021) no.7, 076001
	[arXiv:2106.11659 [hep-ph]].
	
	
	
	%
	\bibitem{Novichkov:2018yse}
	P.~P.~Novichkov, S.~T.~Petcov and M.~Tanimoto,
	Phys.\ Lett.\ B {\bf 793} (2019) 247
	[arXiv:1812.11289 [hep-ph]].
	
	\bibitem{Gui-JunDing:2019wap}
	G.~J.~Ding, S.~F.~King, X.~G.~Liu and J.~N.~Lu,
	JHEP {\bf 1912} (2019) 030
	[arXiv:1910.03460 [hep-ph]].
	
	\bibitem{Okada:2020ukr}
	H.~Okada and M.~Tanimoto,
	Phys. Rev. D \textbf{103} (2021) no.1, 015005
	[arXiv:2009.14242 [hep-ph]].
	
	
	
	
	\bibitem{Novichkov:2021evw}
	P.~P.~Novichkov, J.~T.~Penedo and S.~T.~Petcov,
	JHEP \textbf{04}, 206 (2021)
	[arXiv:2102.07488 [hep-ph]].
	
	\bibitem{Feruglio:2021dte}
	F.~Feruglio, V.~Gherardi, A.~Romanino and A.~Titov,
	JHEP \textbf{05} (2021), 242
	[arXiv:2101.08718 [hep-ph]].
	
	\bibitem{Feruglio:2023bav}
	F.~Feruglio,
	Phys. Rev. Lett. \textbf{130} (2023) no.10, 101801 doi:10.1103/PhysRevLett. 130.101801.
	
	
	\bibitem{Petcov:2022fjf}
	S.~T.~Petcov and M.~Tanimoto,
	[arXiv:2212.13336 [hep-ph]].
	
	\bibitem{Petcov:2023vws}
	S.~T.~Petcov and M.~Tanimoto,
	[arXiv:2306.05730 [hep-ph]].
	
	\bibitem{Kikuchi:2023cap}
	S.~Kikuchi, T.~Kobayashi, K.~Nasu, S.~Takada and H.~Uchida,
	Phys. Rev. D \textbf{107}, no.5, 055014 (2023)
	[arXiv:2301.03737 [hep-ph]].
	
	
	\bibitem{Abe:2023ilq}
	Y.~Abe, T.~Higaki, J.~Kawamurab and T.~Kobayashi,
	[arXiv:2301.07439 [hep-ph]].
	
	
	\bibitem{Kikuchi:2023jap}
	S.~Kikuchi, T.~Kobayashi, K.~Nasu, S.~Takada and H.~Uchida,
	[arXiv:2302.03326 [hep-ph]].
	
	\bibitem{Abe:2023qmr}
	Y.~Abe, T.~Higaki, J.~Kawamura and T.~Kobayashi,
	[arXiv:2302.11183 [hep-ph]].
	
	\bibitem{Abe:2023dvr}
	Y.~Abe, T.~Higaki, J.~Kawamura and T.~Kobayashi,
	[arXiv:2307.01419 [hep-ph]].
	
	\bibitem{Kobayashi:2018wkl}
	T.~Kobayashi, Y.~Shimizu, K.~Takagi, M.~Tanimoto, T.~H.~Tatsuishi and H.~Uchida,
	Phys.\ Lett.\ B {\bf 794} (2019) 114
	[arXiv:1812.11072 [hep-ph]].
	
	
	
	\bibitem{Kobayashi:2019xvz}
	T.~Kobayashi, Y.~Shimizu, K.~Takagi, M.~Tanimoto and T.~H.~Tatsuishi,
	Phys. Rev. D \textbf{100} (2019) no.11, 115045
	[erratum: Phys. Rev. D \textbf{101} (2020) no.3, 039904]
	[arXiv:1909.05139 [hep-ph]].
	
	\bibitem{Asaka:2020tmo}
	T.~Asaka, Y.~Heo and T.~Yoshida,
	Phys. Lett. B \textbf{811} (2020), 135956
	[arXiv:2009.12120 [hep-ph]].
	
	\bibitem{Behera:2020sfe}
	M.~K.~Behera, S.~Mishra, S.~Singirala and R.~Mohanta,
	Phys. Dark Univ. \textbf{36} (2022), 101027
	[arXiv:2007.00545 [hep-ph]].
	
	
	\bibitem{Mishra:2020gxg}
	S.~Mishra,
	[arXiv:2008.02095 [hep-ph]].
	
	
	
	
	
	\bibitem{Okada:2019xqk}
	H.~Okada and Y.~Orikasa,
	Phys. Rev. D \textbf{100}, no.11, 115037 (2019)
	[arXiv:1907.04716 [hep-ph]].
	
	
	\bibitem{Kariyazono:2019ehj}
	Y.~Kariyazono, T.~Kobayashi, S.~Takada, S.~Tamba and H.~Uchida,
	Phys. Rev. D \textbf{100}, no.4, 045014 (2019)
	[arXiv:1904.07546 [hep-th]].
	
	\bibitem{Nomura:2019yft}
	T.~Nomura and H.~Okada,
	Nucl. Phys. B \textbf{966} (2021), 115372
	[arXiv:1906.03927 [hep-ph]].
	
	\bibitem{Okada:2019lzv}
	H.~Okada and Y.~Orikasa,
	arXiv:1908.08409 [hep-ph].
	
	\bibitem{Nomura:2019lnr}
	T.~Nomura, H.~Okada and O.~Popov,
	Phys.\ Lett.\ B {\bf 803} (2020) 135294
	[arXiv:1908.07457 [hep-ph]].
	
	\bibitem{Criado:2019tzk}
	J.~C.~Criado, F.~Feruglio and S.~J.~D.~King,
	JHEP {\bf 2002} (2020) 001
	[arXiv:1908.11867 [hep-ph]].
	
	\bibitem{King:2019vhv}
	S.~F.~King and Y.~L.~Zhou,
	Phys. Rev. D \textbf{101} (2020) no.1, 015001
	[arXiv:1908.02770 [hep-ph]].
	
	
	
	
	
	\bibitem{Zhang:2019ngf}
	D.~Zhang,
	Nucl.\ Phys.\ B {\bf 952} (2020) 114935
	[arXiv:1910.07869 [hep-ph]].
	
	
	\bibitem{Lu:2019vgm}
	J.~N.~Lu, X.~G.~Liu and G.~J.~Ding,
	Phys. Rev. D \textbf{101} (2020) no.11, 115020
	[arXiv:1912.07573 [hep-ph]].
	
	\bibitem{Nomura:2019xsb}
	T.~Nomura, H.~Okada and S.~Patra,
	Nucl. Phys. B \textbf{967} (2021), 115395
	[arXiv:1912.00379 [hep-ph]].
	
	\bibitem{Kobayashi:2019gtp}
	T.~Kobayashi, T.~Nomura and T.~Shimomura,
	Phys. Rev. D \textbf{102} (2020) no.3, 035019
	[arXiv:1912.00637 [hep-ph]].
	
	
	
	
	\bibitem{Wang:2019xbo}
	X.~Wang,
	Nucl. Phys. B \textbf{957} (2020), 115105
	[arXiv:1912.13284 [hep-ph]].
	
	
	\bibitem{King:2020qaj}
	S.~J.~D.~King and S.~F.~King,
	JHEP \textbf{09} (2020), 043
	[arXiv:2002.00969 [hep-ph]].
	
	\bibitem{Abbas:2020qzc}
	M.~Abbas,
	Phys. Rev. D \textbf{103} (2021) no.5, 056016
	[arXiv:2002.01929 [hep-ph]].
	
	
	\bibitem{Okada:2020oxh}
	H.~Okada and Y.~Shoji,
	Phys. Dark Univ. \textbf{31} (2021), 100742
	[arXiv:2003.11396 [hep-ph]].
	
	
	
	\bibitem{Okada:2020dmb}
	H.~Okada and Y.~Shoji,
	Nucl. Phys. B \textbf{961} (2020), 115216
	[arXiv:2003.13219 [hep-ph]].
	
	\bibitem{Ding:2020yen}
	G.~J.~Ding and F.~Feruglio,
	JHEP \textbf{06} (2020), 134
	[arXiv:2003.13448 [hep-ph]].
	
	\bibitem{Nomura:2020opk}
	T.~Nomura and H.~Okada,
	JCAP \textbf{09} (2022), 049
	doi:10.1088/1475-7516/2022/09/049
	[arXiv:2007.04801 [hep-ph]].
	
	
	
	\bibitem{Nomura:2020cog}
	T.~Nomura and H.~Okada,
	arXiv:2007.15459 [hep-ph].
	
	
	\bibitem{Okada:2020rjb}
	H.~Okada and M.~Tanimoto,
	Phys. Dark Univ. \textbf{40} (2023), 101204
	[arXiv:2005.00775 [hep-ph]].
	
	
	\bibitem{deMedeirosVarzielas:2020kji}
	I.~de Medeiros Varzielas, M.~Levy and Y.~L.~Zhou,
	JHEP \textbf{11} (2020), 085
	[arXiv:2008.05329 [hep-ph]].
	
	
	\bibitem{Nagao:2020azf}
	K.~I.~Nagao and H.~Okada,
	JCAP \textbf{05} (2021), 063
	[arXiv:2008.13686 [hep-ph]].
	
	
	
	\bibitem{Nagao:2020snm}
	K.~I.~Nagao and H.~Okada,
	Nucl. Phys. B \textbf{980} (2022), 115841
	[arXiv:2010.03348 [hep-ph]].
	
	
	\bibitem{Abbas:2020vuy}
	M.~Abbas,
	Phys. Atom. Nucl. \textbf{83} (2020) no.5, 764-769.
	
	\bibitem{Kuranaga:2021ujd}
	H.~Kuranaga, H.~Ohki and S.~Uemura,
	JHEP \textbf{07} (2021), 068
	[arXiv:2105.06237 [hep-ph]].
	
	
	\bibitem{Okada:2021aoi}
	H.~Okada and Y.~h.~Qi,
	[arXiv:2109.13779 [hep-ph]].
	
	
	
	\bibitem{Dasgupta:2021ggp}
	A.~Dasgupta, T.~Nomura, H.~Okada, O.~Popov and M.~Tanimoto,
	[arXiv:2111.06898 [hep-ph]].
	
	
	
	\bibitem{Nomura:2021ewm}
	T.~Nomura and H.~Okada,
	Chin. Phys. C \textbf{46} (2022) no.5, 053101
	[arXiv:2109.04157 [hep-ph]].
	
	\bibitem{Nagao:2021rio}
	K.~I.~Nagao and H.~Okada,
	Phys. Dark Univ. \textbf{36} (2022), 101039
	[arXiv:2108.09984 [hep-ph]].
	
	
	
	\bibitem{Nomura:2021yjb}
	T.~Nomura, H.~Okada and Y.~Orikasa,
	Eur. Phys. J. C \textbf{81} (2021) no.10, 947
	[arXiv:2106.12375 [hep-ph]].
	
	
	\bibitem{Nomura:2021aep}
	T.~Nomura and H.~Okada,
	Phys. Rev. D \textbf{105} (2022) no.7, 075010
	[arXiv:2106.10451 [hep-ph]].
	
	
	
	\bibitem{Zhang:2021olk}
	X.~Zhang and S.~Zhou,
	JCAP \textbf{09} (2021), 043
	[arXiv:2106.03433 [hep-ph]].
	
	\bibitem{Wang:2021mkw}
	X.~Wang and S.~Zhou,
	JHEP \textbf{07} (2021), 093
	[arXiv:2102.04358 [hep-ph]].
	
	
	\bibitem{Wang:2020dbp}
	X.~Wang,
	Nucl. Phys. B \textbf{962} (2021), 115247
	[arXiv:2007.05913 [hep-ph]].
	
	
	\bibitem{Ko:2021lpx}
	P.~Ko, T.~Nomura and H.~Okada,
	JHEP \textbf{05} (2022), 098
	[arXiv:2110.10513 [hep-ph]].
	
	\bibitem{Nomura:2021pld}
	T.~Nomura, H.~Okada and Y.~h.~Qi,
	[arXiv:2111.10944 [hep-ph]].
	
	\bibitem{Nomura:2022hxs}
	T.~Nomura and H.~Okada,
	[arXiv:2201.10244 [hep-ph]].
	
	\bibitem{Otsuka:2022rak}
	H.~Otsuka and H.~Okada,
	[arXiv:2202.10089 [hep-ph]].
	
	
	\bibitem{Ding:2021iqp}
	G.~J.~Ding, F.~Feruglio and X.~G.~Liu,
	SciPost Phys. \textbf{10} (2021) no.6, 133
	[arXiv:2102.06716 [hep-ph]].
	
	\bibitem{Charalampous:2021gmf}
	G.~Charalampous, S.~F.~King, G.~K.~Leontaris and Y.~L.~Zhou,
	Phys. Rev. D \textbf{104} (2021) no.11, 115015
	[arXiv:2109.11379 [hep-ph]].
	
	\bibitem{Liu:2021gwa}
	X.~G.~Liu and G.~J.~Ding,
	JHEP \textbf{03} (2022), 123
	[arXiv:2112.14761 [hep-ph]].
	
	
	\bibitem{Novichkov:2022wvg}
	P.~P.~Novichkov, J.~T.~Penedo and S.~T.~Petcov,
	JHEP \textbf{03} (2022), 149
	[arXiv:2201.02020 [hep-ph]].
	
	\bibitem{Kikuchi:2022txy}
	S.~Kikuchi, T.~Kobayashi, H.~Otsuka, M.~Tanimoto, H.~Uchida and K.~Yamamoto,
	Phys. Rev. D \textbf{106} (2022) no.3, 035001
	[arXiv:2201.04505 [hep-ph]].
	
	\bibitem{Behera:2020lpd}
	M.~K.~Behera, S.~Singirala, S.~Mishra and R.~Mohanta,
	J. Phys. G \textbf{49} (2022) no.3, 035002
	[arXiv:2009.01806 [hep-ph]].	
	
	
	
	\bibitem{Ahn:2022ufs}
	Y.~H.~Ahn, S.~K.~Kang, R.~Ramos and M.~Tanimoto,
	Phys. Rev. D \textbf{106} (2022) no.9, 095002
	[arXiv:2205.02796 [hep-ph]].
	
	
	\bibitem{Gunji:2022xig}
	Y.~Gunji, K.~Ishiwata and T.~Yoshida,
	JHEP \textbf{11} (2022), 002
	[arXiv:2208.10086 [hep-ph]].
	\bibitem{Kim:2023jto}
	J.~Kim and H.~Okada,
	[arXiv:2302.09747 [hep-ph]].
	
	
	
	\bibitem{Nomura:2023dgk}
	T.~Nomura, H.~Okada and Y.~Shoji,
	PTEP \textbf{2023} (2023) no.2, 023B04.
	
	
	
	
	\bibitem{Kang:2022psa}
	D.~W.~Kang, J.~Kim, T.~Nomura and H.~Okada,
	JHEP \textbf{07} (2022), 050
	[arXiv:2205.08269 [hep-ph]].
	
	
	
	\bibitem{Abe:2023ylh}
	Y.~Abe, T.~Higaki, F.~Kaneko, T.~Kobayashi and H.~Otsuka,
	[arXiv:2303.02947 [hep-ph]].
	
	
	
	
	
	\bibitem{deMedeirosVarzielas:2022fbw}
	I.~de Medeiros Varzielas, S.~F.~King and M.~Levy,
	JHEP \textbf{02} (2023), 143
	[arXiv:2211.00654 [hep-ph]].
	
	
	\bibitem{Ding:2023ynd}
	G.~J.~Ding, S.~F.~King, C.~C.~Li, X.~G.~Liu and J.~N.~Lu,
	[arXiv:2303.02071 [hep-ph]].
	
	\bibitem{deAnda:2023udh}
	F.~J.~de Anda and S.~F.~King,
	[arXiv:2304.05958 [hep-ph]].
	
	\bibitem{Nomura:2023kwz}
	T.~Nomura and H.~Okada,
	[arXiv:2304.13361 [hep-ph]].
	
	\bibitem{Ahn:2023iqa}
	Y.~H.~Ahn and S.~K.~Kang,
	[arXiv:2306.14467 [hep-ph]].
	
	
	\bibitem{Kobayashi:2016ovu}
	T.~Kobayashi, S.~Nagamoto and S.~Uemura,
	PTEP \textbf{2017}, no.2, 023B02 (2017)
	[arXiv:1608.06129 [hep-th]].
	
	
	\bibitem{deMedeirosVarzielas:2019cyj}
	I.~de Medeiros Varzielas, S.~F.~King and Y.~L.~Zhou,
	Phys.\ Rev.\ D {\bf 101} (2020) no.5,  055033
	[arXiv:1906.02208 [hep-ph]].
	
	
	\bibitem{Ishiguro:2020tmo}
	K.~Ishiguro, T.~Kobayashi and H.~Otsuka,
	JHEP \textbf{03}, 161 (2021)
	[arXiv:2011.09154 [hep-ph]].
	
	
	
	
	
	\bibitem{Abe:2020vmv}
	H.~Abe, T.~Kobayashi, S.~Uemura and J.~Yamamoto,
	Phys. Rev. D \textbf{102}, no.4, 045005 (2020)
	[arXiv:2003.03512 [hep-th]].
	
	
	
	
	
	
	\bibitem{Kikuchi:2021yog}
	S.~Kikuchi, T.~Kobayashi, Y.~Ogawa and H.~Uchida,
	PTEP \textbf{2022}, no.3, 033B10 (2022)
	[arXiv:2112.01680 [hep-ph]].
	
	
	
	
	
	
	
	\bibitem{Nilles:2020nnc}
	H.~P.~Nilles, S.~Ramos-\'Sanchez and P.~K.~S.~Vaudrevange,
	JHEP \textbf{02} (2020), 045
	[arXiv:2001.01736 [hep-ph]].
	
	\bibitem{Nilles:2020kgo}
	H.~P.~Nilles, S.~Ramos-S\'anchez and P.~K.~S.~Vaudrevange,
	Nucl. Phys. B \textbf{957} (2020), 115098
	[arXiv:2004.05200 [hep-ph]].
	
	\bibitem{Nilles:2020tdp}
	H.~P.~Nilles, S.~Ramos\textendash{}S\'anchez and P.~K.~S.~Vaudrevange,
	Phys. Lett. B \textbf{808} (2020), 135615
	[arXiv:2006.03059 [hep-th]].
	
	
	
	
	
	\bibitem{Baur:2020yjl}
	A.~Baur, M.~Kade, H.~P.~Nilles, S.~Ramos-Sanchez and P.~K.~S.~Vaudrevange,
	Phys. Lett. B \textbf{816} (2021), 136176
	[arXiv:2012.09586 [hep-th]].
	
	
	
	\bibitem{Ding:2020zxw}
	G.~J.~Ding, F.~Feruglio and X.~G.~Liu,
	JHEP \textbf{01} (2021), 037
	[arXiv:2010.07952 [hep-th]].
	
	
	
	
	\bibitem{Baur:2021mtl}
	A.~Baur, M.~Kade, H.~P.~Nilles, S.~Ramos-Sanchez and P.~K.~S.~Vaudrevange,
	JHEP \textbf{06} (2021), 110
	[arXiv:2104.03981 [hep-th]].
	
	\bibitem{Nilles:2021ouu}
	H.~P.~Nilles, S.~Ramos-Sanchez and P.~K.~S.~Vaudrevange,
	[arXiv:2105.02984 [hep-th]].
	
	\bibitem{Nilles:2021glx}
	H.~P.~Nilles, S.~Ramos-Sanchez, A.~Trautner and P.~K.~S.~Vaudrevange,
	Nucl. Phys. B \textbf{971} (2021), 115534
	[arXiv:2105.08078 [hep-th]].
	
	\bibitem{Baur:2021bly}
	A.~Baur, H.~P.~Nilles, S.~Ramos-Sanchez, A.~Trautner and P.~K.~S.~Vaudrevange,
	Phys. Rev. D \textbf{105} (2022) no.5, 055018
	[arXiv:2112.06940 [hep-th]].
	
	\bibitem{Baur:2022hma}
	A.~Baur, H.~P.~Nilles, S.~Ramos-Sanchez, A.~Trautner and P.~K.~S.~Vaudrevange,
	JHEP \textbf{09} (2022), 224
	[arXiv:2207.10677 [hep-ph]].
	
	
	
	
	\bibitem{Kikuchi:2023clx}
	S.~Kikuchi, T.~Kobayashi, K.~Nasu, H.~Otsuka, S.~Takada and H.~Uchida,
	JHEP \textbf{04}, 003 (2023)
	[arXiv:2301.10356 [hep-th]].
	
	
	
	\bibitem{Feruglio:2023mii}
	F.~Feruglio,
	JHEP \textbf{03} (2023), 236
	[arXiv:2302.11580 [hep-ph]].
	
	
	\bibitem{Knapp-Perez:2023nty}
	V.~Knapp-Perez, X.~G.~Liu, H.~P.~Nilles, S.~Ramos-Sanchez and M.~Ratz,
	[arXiv:2304.14437 [hep-th]].
	
	
	 
	
	
	\bibitem{Abe:2008fi}
	H.~Abe, T.~Kobayashi and H.~Ohki,
	JHEP \textbf{09} (2008), 043
	[arXiv:0806.4748 [hep-th]].
	
	
	\bibitem{Abe:2013bca} 
	T.~H.~Abe, Y.~Fujimoto, T.~Kobayashi, T.~Miura, K.~Nishiwaki and M.~Sakamoto,
	JHEP {\bf 1401}, 065 (2014)
	[arXiv:1309.4925 [hep-th]].
	
	
	
	
	\bibitem{Abe:2017gye}
	H.~Abe, T.~Kobayashi, K.~Sumita and S.~Uemura,
	Phys. Rev. D \textbf{96}, no.2, 026019 (2017)
	[arXiv:1703.03402 [hep-th]].
	
	
	
	\bibitem{Abe:2012fj}
	H.~Abe, T.~Kobayashi, H.~Ohki, A.~Oikawa and K.~Sumita,
	Nucl. Phys. B \textbf{870}, 30-54 (2013)
	[arXiv:1211.4317 [hep-ph]].  
	
	
	\bibitem{Abe:2014vza}
	H.~Abe, T.~Kobayashi, K.~Sumita and Y.~Tatsuta,
	Phys. Rev. D \textbf{90}, no.10, 105006 (2014)
	[arXiv:1405.5012 [hep-ph]].
	
	
	\bibitem{Fujimoto:2016zjs}
	Y.~Fujimoto, T.~Kobayashi, K.~Nishiwaki, M.~Sakamoto and Y.~Tatsuta,
	Phys. Rev. D \textbf{94}, no.3, 035031 (2016)
	[arXiv:1605.00140 [hep-ph]].
	
	\bibitem{Buchmuller:2017vho}
	W.~Buchmuller and J.~Schweizer,
	Phys. Rev. D \textbf{95}, no.7, 075024 (2017)
	[arXiv:1701.06935 [hep-ph]].
	
	\bibitem{Buchmuller:2017vut}
	W.~Buchmuller and K.~M.~Patel,
	Phys. Rev. D \textbf{97}, no.7, 075019 (2018)
	[arXiv:1712.06862 [hep-ph]].
	
	
	
	\bibitem{Hoshiya:2022qvr}
	K.~Hoshiya, S.~Kikuchi, T.~Kobayashi and H.~Uchida,
	Phys. Rev. D \textbf{106} (2022) no.11, 115003
	[arXiv:2209.07249 [hep-ph]].
	
	
	\bibitem{Kikuchi:2022svo}
	S.~Kikuchi, T.~Kobayashi, M.~Tanimoto and H.~Uchida,
	[arXiv:2207.04609 [hep-ph]].
	
	
	\bibitem{Fukugita:2003tn}
	M.~Fukugita, M.~Tanimoto and T.~Yanagida,
	Phys. Lett. B \textbf{562} (2003), 273-278
	doi:10.1016/S0370-2693(03)00568-9
	[arXiv:hep-ph/0303177 [hep-ph]].
	
	\bibitem{Xing:2004xu}
	Z.~z.~Xing and S.~Zhou,
	Phys. Lett. B \textbf{593} (2004), 156-164
	[arXiv:hep-ph/0403261 [hep-ph]].
	
	\bibitem{Obara:2006ny}
	M.~Obara and Z.~z.~Xing,
	Phys. Lett. B \textbf{644} (2007), 136-146
	doi:10.1016/j.physletb. 2006.11.010 [arXiv:hep-ph/0608280 [hep-ph]].
	
	\bibitem{Fukugita:2012jr}
	M.~Fukugita, Y.~Shimizu, M.~Tanimoto and T.~T.~Yanagida,
	Phys. Lett. B \textbf{716} (2012), 294-297
	doi:10.1016/j.physletb.2012.06.049
	[arXiv:1204.2389 [hep-ph]].
	
	\bibitem{Fukugita:2016hzf}
	M.~Fukugita, Y.~Kaneta, Y.~Shimizu, M.~Tanimoto and T.~T.~Yanagida,
	Phys. Lett. B \textbf{764} (2017), 163-166
	doi:10.1016/j.physletb.2016.11.024
	[arXiv:1609.01864 [hep-ph]].
	
	
	\bibitem{Fritzsch:2012rg}
	H.~Fritzsch and S.~Zhou,
	Phys. Lett. B \textbf{718} (2013), 1457-1464
	[arXiv:1212.0411 [hep-ph]].
	
	
	
	
	
	\bibitem{Gukov:1999ya}
	S.~Gukov, C.~Vafa and E.~Witten,
	Nucl. Phys. B \textbf{584}, 69-108 (2000)
	[erratum: Nucl. Phys. B \textbf{608}, 477-478 (2001)]
	[arXiv:hep-th/9906070 [hep-th]].
	
	
	
	

	
	\bibitem{Ishiguro:2022pde}
	K.~Ishiguro, H.~Okada and H.~Otsuka,
	JHEP \textbf{09} (2022), 072
	[arXiv:2206.04313 [hep-ph]].
	
	

	
	
	
	\bibitem{Cvetic:2003ch}
	M.~Cvetic and I.~Papadimitriou,
	Phys. Rev. D \textbf{68}, 046001 (2003)
	[erratum: Phys. Rev. D \textbf{70}, 029903 (2004)]
	[arXiv:hep-th/0303083 [hep-th]].
	
	\bibitem{Abel:2003vv}
	S.~A.~Abel and A.~W.~Owen,
	Nucl. Phys. B \textbf{663}, 197-214 (2003)
	[arXiv:hep-th/0303124 [hep-th]].
	
	
	\bibitem{Abel:2003yx}
	S.~A.~Abel and A.~W.~Owen,
	Nucl. Phys. B \textbf{682}, 183-216 (2004)
	[arXiv:hep-th/0310257 [hep-th]].
	
	
	
	
	\bibitem{Cremades:2004wa}
	D.~Cremades, L.~E.~Ibanez and F.~Marchesano,
	JHEP \textbf{05} (2004), 079
	[arXiv:hep-th/0404229 [hep-th]].
	
	
	\bibitem{Abe:2009dr}
	H.~Abe, K.~S.~Choi, T.~Kobayashi and H.~Ohki,
	JHEP \textbf{06}, 080 (2009)
	[arXiv:0903.3800 [hep-th]].
	
	
	\bibitem{Hamidi:1986vh}
	S.~Hamidi and C.~Vafa,
	Nucl. Phys. B \textbf{279}, 465-513 (1987).
	
	\bibitem{Dixon:1986qv}
	L.~J.~Dixon, D.~Friedan, E.~J.~Martinec and S.~H.~Shenker,
	Nucl. Phys. B \textbf{282}, 13-73 (1987).
	
	\bibitem{Atick:1987kd}
	J.~J.~Atick, L.~J.~Dixon, P.~A.~Griffin and D.~Nemeschansky,
	Nucl. Phys. B \textbf{298}, 1-35 (1988).
	
	
	\bibitem{Burwick:1990tu}
	T.~T.~Burwick, R.~K.~Kaiser and H.~F.~Muller,
	Nucl. Phys. B \textbf{355}, 689-711 (1991).
	
	
	\bibitem{Stieberger:1992bj}
	S.~Stieberger, D.~Jungnickel, J.~Lauer and M.~Spalinski,
	Mod. Phys. Lett. A \textbf{7}, 3059-3070 (1992)
	[arXiv:hep-th/9204037 [hep-th]].
	
	
	\bibitem{Choi:2007nb}
	K.~S.~Choi and T.~Kobayashi,
	Nucl. Phys. B \textbf{797}, 295-321 (2008)
	[arXiv:0711.4894 [hep-th]].
	
	
	
	
	
	
	
\end{thebibliography}

\end{document}